\documentclass{elsart}
\usepackage{epsfig}
\usepackage{epstopdf}

\usepackage{graphics,epsfig}
\usepackage{graphicx}
\usepackage{bm}
\usepackage{amsmath}
\usepackage{mathrsfs}

\newcommand{\be}{\begin{equation}}
\newcommand{\ee}{\end{equation}}
\newcommand{\beq}{\begin{eqnarray}}
\newcommand{\eeq}{\end{eqnarray}}

\def\aprle{\buildrel < \over {_{\sim}}}
\def\aprge{\buildrel > \over {_{\sim}}}

\def\t13{\mathrel{{\theta_{13}}}}
\def\y12{\mathrel{{\tan^2 \theta_{12}}}}
\def\c2{\mathrel{{\chi^2 }}}

\newcommand{\ceci}[1]{{}}

\begin{document}
\begin{frontmatter}

\title{Oscillations of very low energy atmospheric neutrinos 
}
\author{Orlando L. G. Peres}
\address{The Abdus Salam International Centre for Theoretical Physics,
I-34100 Trieste, Italy}
\address{Instituto de F\'\i sica Gleb Wataghin - UNICAMP, 
  13083-970 Campinas SP, Brazil}

\author{A. Yu. Smirnov}
\address{The Abdus Salam International Centre for Theoretical Physics,
I-34100 Trieste, Italy}
\address{Institute for Nuclear Research of Russian Academy
of Sciences, Moscow 117312, Russia}

\date{\today}

 \begin{abstract}

There are several new features in production, oscillations and detection 
of the atmospheric neutrinos of low energies, $E \aprle 100$ MeV. 
The flavor ratio, $r$, of muon to electron neutrino 
fluxes is substantially smaller than 2 and decreases with energy,  
significant part of events is due to the decay of invisible muons at 
rest, etc. Oscillations in two-layer medium (atmosphere - earth) should 
be taken into account. We derive analytical and semi-analytical 
expressions  
for the oscillation  probabilities of these ``sub-subGeV'' neutrinos. 
The energy spectra of the $e-$like events in water cherenkov 
detectors are computed and dependence of the spectra  on 
the 2-3 mixing angle,  $\theta_{23}$, the  1-3 mixing  and CP-violation 
phase are studied.  We find that variations of  $\theta_{23}$ in the 
presently  allowed  region change the number of $e-$like events by about 
$15 - 20\%$ as well as to distortion of the energy spectrum. 
The 1-3  mixing and CP-violation can lead  to 
$\sim 10\%$ effects. Detailed study of the sub-subGeV neutrinos 
will be possible in future Megaton-scale detectors. 
 
 \end{abstract}

\begin{keyword}
Atmospheric Neutrinos, Oscillations 
\end{keyword}
\end{frontmatter}

\section{Introduction}

Studies of the atmospheric neutrinos have been performed  
mainly at energies $E \aprge (0.1- 0.2)$ GeV.  
There are only few results on neutrinos of lower energies:  
$E = (10 - 100)$ MeV. 
The fluxes of these neutrinos \cite{zatsepin:1962} have been computed 
recently by several different groups \cite{Gaisser:1983vc,Gaisser:1988ar},  
\cite{Honda:1995hz,hondar:1990ar}, \cite{Battistoni:2005pd}, 
and there are some differences 
of the results of  computations.   

The $e-$like events induced by the low energy
atmospheric  neutrinos have been detected by SuperKamiokande 
\cite{Malek:2002ns,Malek:2003ki,skiida}.    
About $88 \pm 12$ events were produced by interactions of the atmospheric 
$\nu_e$ and $\bar{\nu}_e$ directly, and 
$174 \pm 16$ events originated from decays of invisible muons. 
In turn, these muons are generated by the 
$\nu_\mu$-flux with typical energies (150 - 250) MeV \cite{Malek:2003ki}. 
LSD  put only an upper bound on the $\bar{\nu}_e-$flux:  
$F_{\bar{e}} < 5 \cdot 10^{4} {\rm cm}^{-2}{\rm s}^{-1}$  \cite{lsd}
which is about 5 times higher than the expected value for the energy 
range $12 < E < 26$ MeV.   

As far as the oscillation effects are concerned, only  
the vacuum $\nu_{\mu} - \nu_{\tau}$ oscillations have  
been taken into account in \cite{Malek:2002ns,Malek:2003ki,skiida}.
Also oscillations in two layer medium with constant densities 
relevant for low energy neutrinos have been considered \cite{2layer}.

The low energy atmospheric neutrinos were discussed, 
as a background for detection of the 
relic supernova neutrinos \cite{Fogli:2004gy} 
as well as future SN neutrino bursts~\cite{Fogli:2004ff}.

In this paper we perform detailed study of the oscillations 
of the ``sub-sub-GeV'' atmospheric neutrinos in the complete 
$3\nu$-context.  
The range below 100 MeV offers rather rich oscillation phenomenology.  
Although presently the number of detected events is 
small,  in future,  new large  scale experiments 
can accumulated large enough statistics to extract new interesting 
information.

The paper is organized as follows. In sect. 2 we summarize 
properties of the neutrino fluxes at low energies, as well as 
parameters of the neutrino trajectories.  
We present relevant oscillation probabilities inside the Earth 
in sect. 3. The probabilities of $3\nu$-oscillations in  two layer 
medium  (the atmosphere and the Earth) are derived in sect. 4. 
In sect. 5  we  consider averaging and integration 
of the probabilities over the  angular variables. In sect. 6.  
we present the $\nu_e$, $\nu_\mu$ atmospheric neutrino 
fluxes at a detector. We then compute the numbers of $e-$like events 
as functions of energy, in water Cherenkov detectors induced by the 
direct $\nu_e$, $\bar{\nu}_e$ interactions (sect. 7) and via the 
invisible muon decays (sect. 8). We study dependence of observables 
on the oscillation parameters. 
In sect. 9 we present the energy spectra of $e-$like events in 
the  megaton-scale detectors. 
Discussion and conclusions follow in sect. 10.    
In Appendices A, B, C we present analytic and semi-analytic
formulas for the oscillation probabilities.

\section{Atmospheric $\nu$ fluxes and trajectories}

\begin{figure}[htbp]
\includegraphics[scale=0.7]{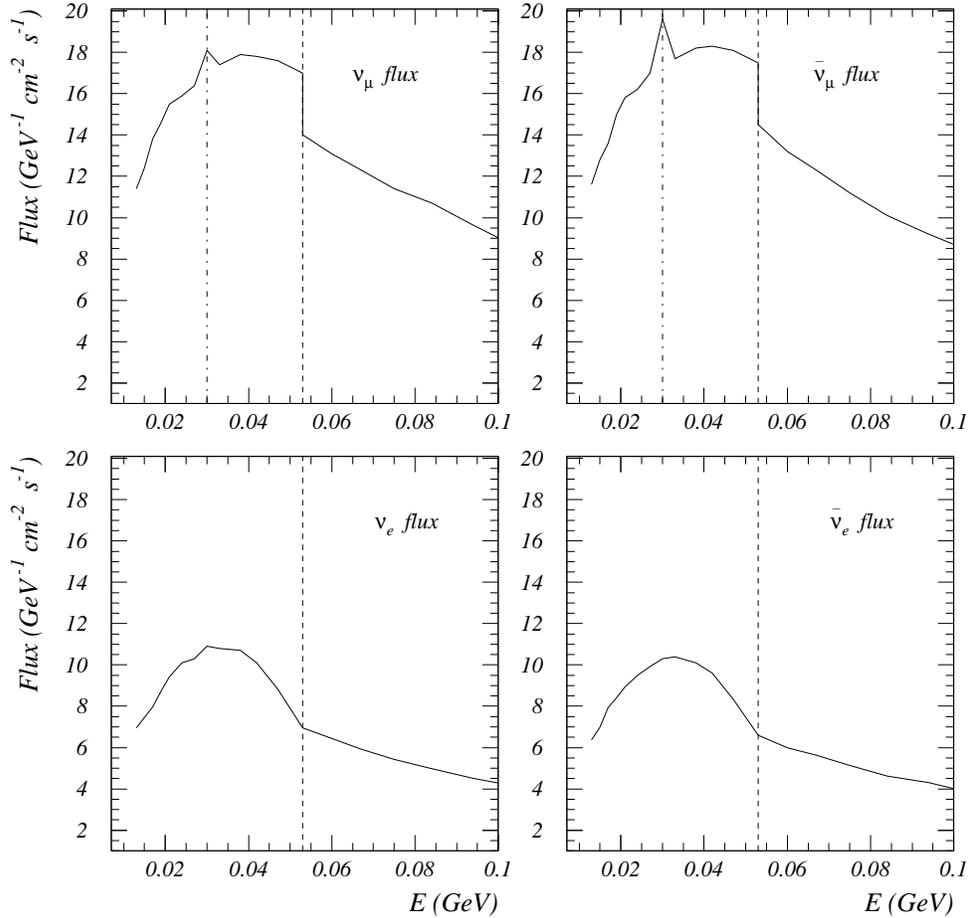}
\caption{The fluxes of $\nu_{\mu}$, $\bar{\nu}_{\mu}$, $\nu_{e}$ and
$\bar{\nu}_{e}$ neutrinos as functions of neutrino energy 
from  \cite{Battistoni:2005pd}.  
The dashed vertical lines show the end point of the 
neutrino energy spectrum from the muon decay at rest, $E^{\mu}_\nu$, and 
the neutrino energy from the $\pi-$decay at rest, $E^{\pi}_\nu$.}
\label{f1}
\end{figure}

Let us summarize properties of the neutrino fluxes with 
$E \aprle 0.1$ GeV.  
In fig.~\ref{f1} we show the produced (without oscillations)  
fluxes  of the electron neutrinos,
$F_e^0$, electron antineutrinos, $F_{\bar e}^0$, 
muon neutrinos, $F_\mu^0$, and muon antineutrinos,  $F_{\bar \mu}^0$, 
as functions of the neutrino energy.  
The lines have been drawn using results of computations 
in ref.  \cite{Battistoni:2005pd}.  
The fluxes are averaged over the zenith and azimuthal angles. 
They depend on period of solar activity 
which influences the fluxes of primary cosmic rays. 
Shown in the figure are  the average values of neutrino fluxes  
during periods of  the  maximum and minimum of solar activity. 

The low energy neutrino fluxes are formed  mainly in the decays  
of pions and muons in flight and at rest. 
The K-meson decays contribute less than $0.05\%$.  
The key energy scales (indicated by the dashed lines in the figures) 
are  the energy of muon neutrinos from the pion decay at rest,  
$E^{\pi}_{\nu} \approx 30$ MeV, and the end point of   
neutrino spectrum from the muon decay, $E^{\mu}_{\nu} = 53$ MeV. 

Properties of the  spectra can be summarized as follows. 

\begin{enumerate}

\item For $E > E^\mu_\nu = 53$ MeV 
the $\nu_e-$ and $\bar{\nu}_e-$ spectra are formed 
in muon decays in flight. 

\item The bumps  in $\nu_{e}-$ and  $\bar{\nu}_{e}-$  spectra 
below $E^\mu_\nu = 53$ MeV  are due to the muon decay at rest. 
This contribution composes about 1/3  of the total flux at these 
energies.    

\item The  neutrino flux is slightly  $(5- 10 \%)$ larger than the 
antineutrino flux:  $F_e^0 > F_{\bar e}^0$. 
The reason is that $\nu_{e}$'s originate from the chain of reactions  
$\pi^+\to \mu^+ \nu_{\mu}$, 
$\mu^+ \to e^+ \nu_e \bar{\nu}_{\mu}$, whereas  
$\bar{\nu}_{e}$ -  from
the conjugate reactions. Since the original cosmic rays are 
protons and nuclei,  they overproduce $\pi^+$ in comparison with $\pi^-$,
and consequently, the $\pi^+$ chain is more abundant.

\item For $E > 53$ MeV  the muon (anti)neutrino spectra are formed by the 
pion and muon decays in flight. Since both the 
$\pi^+ -$ and $\pi^- - $decay chains produce equal number of 
$\nu_{\mu}$ and $\bar{\nu}_{\mu}$, the corresponding fluxes are 
approximately equal. The bump in the spectrum  below $E = 53$ MeV  with 
sharp edge at $E = 53$ MeV, is due to muon  decay at rest. The peak at 
$E^{\pi}_{\nu} = 30$ MeV originates from  the pion decay at rest. 
Below $30$ MeV the main contribution to the $\nu_{\mu}-$flux is from the 
muon decay, and about $38\%$ of the flux is generated 
by the pion decay in flight with neutrinos emitted in non-forward 
directions.  

\item Below $E^{\mu}_{\nu}$, the  $\bar{\nu}_{\mu}-$flux 
is slightly larger than the $\nu_{\mu}-$ flux:
$F_{\bar \mu}^0 = 1.05 F_\mu^0$. 
The difference originates from the muon decay at rest and has 
the same reason as larger flux of $\nu_e$:    
$\bar{\nu}_{\mu}$ comes from the chain 
$\pi^+ \rightarrow \mu^+ \rightarrow \bar{\nu}_{\mu}$. 

\item According to \cite{Battistoni:2005pd} the $\pi-$decay peak for  
$\bar{\nu}_{\mu}$ is larger than the ${\nu}_{\mu}-$peak.  

\end{enumerate}


The flavor ratios  
\be 
r(E, \Theta_\nu) \equiv \frac{F_{\mu}^0(E, \Theta_\nu)}{ F_e^0(E,
\Theta_\nu)}, ~~~~
\bar{r}(E, \Theta_\nu) \equiv 
\frac{F_{\bar \mu}^0(E, \Theta_\nu)}{F_{\bar e}^0(E,\Theta_\nu)}, 
\label{f-ratio}
\ee
play the key role in oscillations. Here $\Theta_\nu$ is the zenith angle 
of neutrino trajectory.   
In what follow we will discuss the ratios averaged over the 
zenith and azimuthal angles. 
As follows from  fig.~\ref{f1},  in the range $E > E^{\mu}_{\nu}$
the ratios equal $r \approx 2.0$ and $\bar{r} \approx 2.2$.  
Below $E^{\mu}_{\nu}$ the ratios decrease with energy.  
For instance, in maximum, $E \sim (35 - 40)$ MeV, one has   
$r \approx 1.72$ and $\bar{r} \approx 1.87$; 
for $E = 21$ MeV we obtain $r \approx 1.65$ and $\bar{r} \approx 1.77$. 
This is important difference from the case of higher energy atmospheric 
neutrinos, where $r \sim 2$ and $\bar{r} \sim 2.2$. 

\begin{figure}[h]
  \includegraphics[scale=0.5]{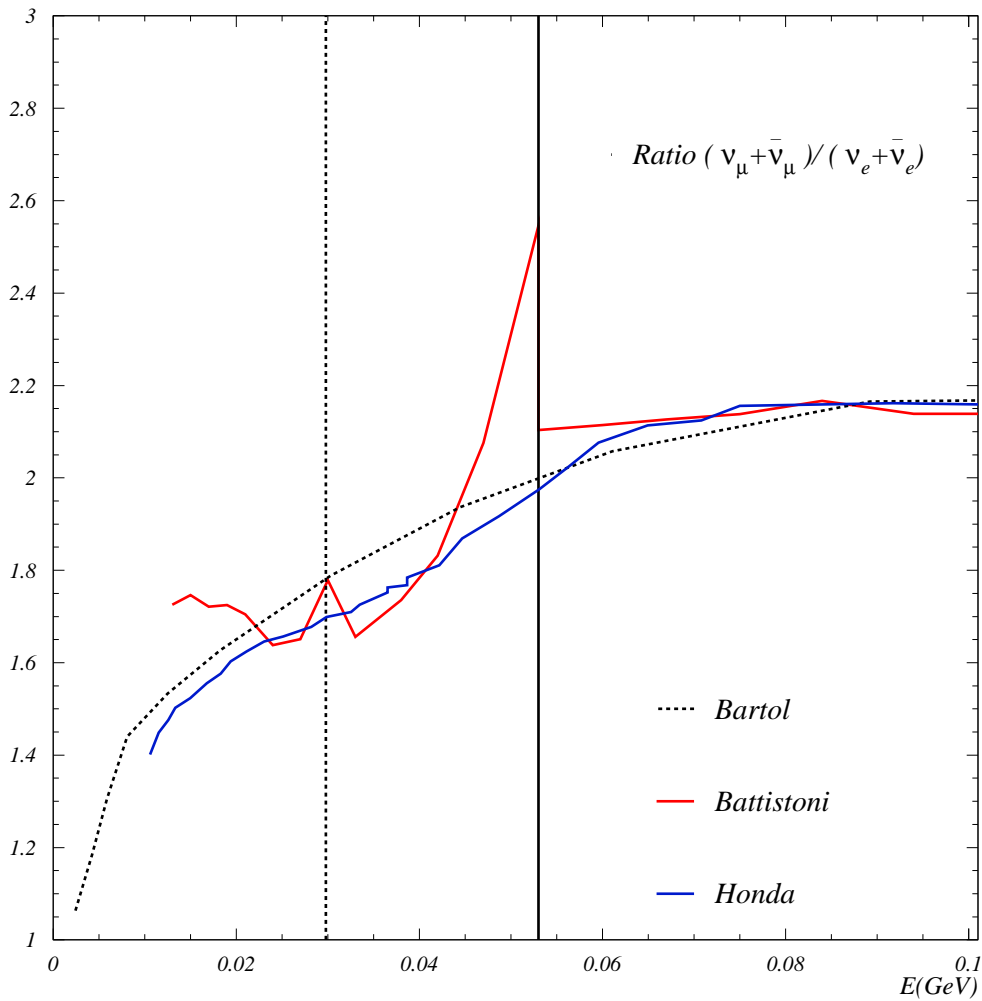}
\caption{The flavor ratio (\ref{rsum}) 
obtained from results of computations of different groups: dashed line  
\cite{Gaisser:1988ar} ,  red line
\cite{Battistoni:2005pd}, blue line  
\cite{hondar:1990ar}. 
The vertical lines indicate the neutrino energy   
from the pion decay, $E^{\pi}_\nu$ (dashed), and 
the end-point of muon decay spectrum, $E^{\mu}_\nu$ (solid). }
\label{compare3}
\end{figure}

In fig.~\ref{compare3}, we show the flavor ratio 
for the sum of neutrino and antineutrino fluxes: 
\be
\frac{F_{\mu}^0 + F^0_{\bar \mu}}{F^0_{e} + F^0_{\bar e}}
\label{rsum}
\ee
obtained from different computations. 
The  ratio changes from  2 to almost 1.5, when  the
neutrino energy decreases from 100 to 10 MeV. 
Sharp jump of the ratio at 53 MeV is due to  
the muon decay spectrum which has maximum at the end point.

\begin{figure}[htbp]
\includegraphics[scale=0.5]{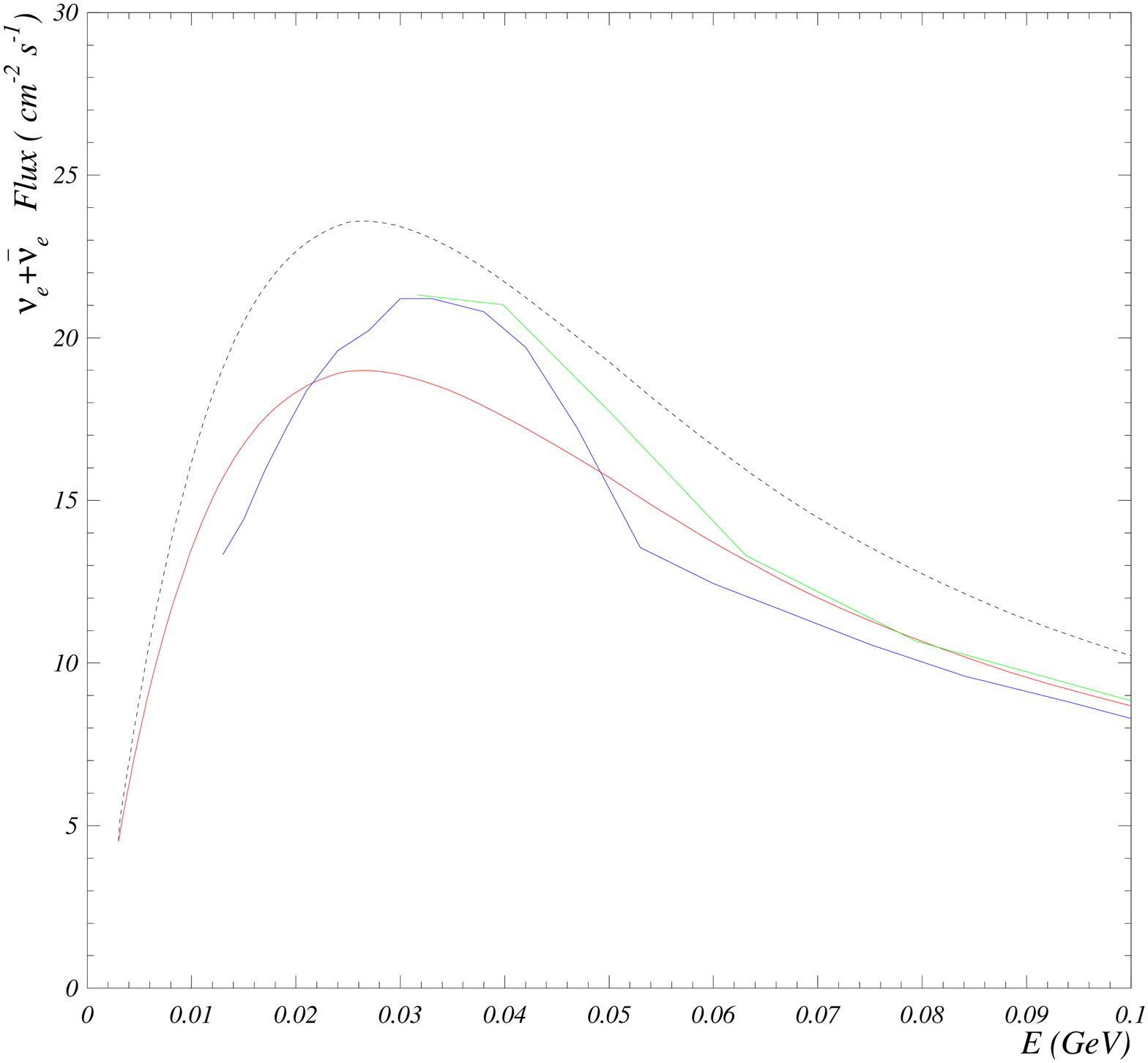}
\caption{The  ($\nu_{e}+\bar{\nu}_{e}$)- neutrino flux 
as function of energy for  
maximum  (solid red)
and for minimum (solid dashed) of the solar activity  
from \cite{Gaisser:1988ar}. 
Show are also the averaged over the solar activity cycle fluxes from   
\cite{hondar:1990ar} - Honda (green) and \cite{Battistoni:2005pd} - FLUKA  (blue).}
\label{f0}
\end{figure}

There are differences between neutrino spectra 
presented by  different groups. In fig.~\ref{f0}
we compare  the sum of $\nu_e$ and $\bar{\nu}_e$- fluxes from three 
available computations. Notice that above 20 - 30 MeV 
the difference of spectra 
is about ($10 - 15 \% $) and the shapes of spectra are rather similar. 
To have an idea about the solar activity effect we show in fig.~\ref{f0} 
the fluxes in maximum and minimum of the activity 
computed by  Bartol  group \cite{Gaisser:1988ar}.  The difference is 
$(15 - 20) \%$. In what follows we  will use the averaged over 
the solar cycle fluxes.

In contrast to high energies, 
for low energy neutrinos one needs to take into account  
the oscillations driven by the 1-2 mixing in atmosphere. 
Indeed, the oscillation length is given by 
\be
l_\nu \equiv \frac{4 \pi E}{\Delta m^2_{21}} =  
10^3 {\rm km}  \left(\frac{E}{30 {\rm MeV}}\right)
\ee
which is comparable to the length of trajectory 
in the horizontal direction. 

The total length of neutrino trajectory from a production point to 
a detector, $L$, is given by 
\be
L^2 = 2R \left[R \cos^2 \Theta_\nu + h - 
\cos\Theta_\nu \sqrt{R^2 \cos^2\Theta_\nu + 2 Rh + h^2} +\frac{h^2}{2R}
\right], 
\label{totlen}
\ee 
where $R$ is the radius of the Earth, $h\sim 20$ km  is the height 
in the atmosphere where neutrinos are produced, $\Theta_\nu$ is the 
zenith  angle. We neglect here a depth of detector below the 
surface of the Earth.  
 
Above the horizon, $\cos \Theta_\nu > 0$,  trajectories are in the 
atmosphere only. In the horizontal direction, $\cos \Theta_\nu = 0$:
$L = \sqrt{2Rh + h^2} \approx \sqrt{2Rh}$. 
For trajectories below the horizon $\cos \Theta_\nu < 0$  
the length of trajectory inside the matter of the Earth equals 
\be
L_m = - 2R \cos\Theta_\nu,  
\ee
and for these trajectories the length in the atmosphere: 
\be
L_A(\cos \Theta_\nu) = L(\cos\Theta_\nu) - L_m(\cos \Theta_\nu) = 
L(- \cos \Theta_\nu).
\ee     
For trajectories not very close to the horizon we have
$L_A \approx h/ \cos \Theta_\nu$ which corresponds to the flat 
atmosphere and $L \approx 2 \sqrt{r(R \cos^2 \Theta_\nu + h)}$.  
The length of trajectory in mater and in atmosphere 
become comparable, $L_A = L_m \sim 500$ km 
at $|\cos \Theta_\nu| \approx \sqrt{h/2R} = 0.04$.

The half-phases of oscillations in the atmosphere and the 
Earth equal  
\be
\phi = \pi \frac{L_A}{l_\nu}, ~~~~ \phi^m = \pi \frac{L_m}{l_m},
\label{phase-m}
\ee    
with  $l_\nu$ and $l_m$ being the oscillation lengths in 
vacuum and in matter correspondingly.  
Notice that at low energies the phase  in the atmosphere can not be 
neglected even for trajectories not very close to 
the horizon: for instance, for $E = 30$ MeV 
and $\cos \Theta_\nu = -0.3$  we obtain $\sin \phi = 0.2$. 
For $E = 60$ MeV, we have $\sin \phi = 0.1$. 
So, in general at  low energies  the  
vacuum oscillation phase can not be treated  
as a small parameter. 
The oscillations in  atmosphere can be neglected 
for the muon neutrinos producing  the invisible muons. 

\section{Oscillation effects inside the Earth} 

Let us first consider the main features of oscillations 
at low energies inside the Earth. 
The resonance energy equals 
\be
E_R = \frac{\Delta m^2_{21} \cos 2\theta_{12}}{2V_e} = 96.4\;{\rm MeV} 
\left( \frac{\Delta m^2_{21}}{7.3\cdot  10^{-5}{\rm eV}^2 } \right)
\left( \frac{2.0{\rm g}/{\rm cm}^3}{Y_e\rho} \right)
\left(\frac{\cos 2\theta_{12}}{0.424}\right) .
\ee
So, the energy interval $E \aprle 0.1$ GeV is at and below the 1-2 
resonance. 
It is much below the 1-3 resonance: 
$E/E_R^{(13)} \aprle 0.015$.  
As a consequence,

\begin{itemize}

\item
Matter effect on the 1-3 mixing is very small and 
can be neglected in the first approximation. 

\item
The third eigenstate approximately coincides with 
the third mass eigenstate $\nu_{3m} \approx \nu_3$. 

\item
The state $\nu_3$ decouples from dynamics and evolves independently. 
The two other states form $2\nu-$mixing system, 
and so the problem is reduced to $2\nu-$ problem. 
   
\item
Oscillations driven by the large mass split, $\Delta m_{31}^2$,  
are averaged out in probability due to integration over the angular 
variables 
and energy.   

\end{itemize}
   
In what follows we will quantify this picture and derive the relevant 
oscillation probabilities. 
Evolution of the neutrino flavor states, $\nu_f \equiv (\nu_e,
\nu_{\mu}, \nu_{\tau})^T$,  is described by the equation 
\be
i \frac{d \nu_f}{dt} =
\left( \frac{U M^2 U^\dagger}{2 E} + \hat{V} \right) \nu_f, 
\label{evolution}
\ee 
where the PMNS mixing matrix defined through $\nu_f = U \nu_{mass}$ 
can be parameterized as $U = U_{23} \Gamma_\delta U_{13} U_{12}$. 
Here $\Gamma_\delta \equiv {\rm diag}(1, 1, e^{i\delta})$, 
$\hat{V} \equiv diag(V, 0, 0)$, and $V = \sqrt{2} G_F n_e$, 
and $U_{ij}$ is the  rotation in the $ij-$plane onto the angle 
$\theta_{ij}$.

Consider new basis of states, 
$\nu^{\prime} = ({\nu}_e^{\prime}, 
{\nu}_{\mu}^{\prime}, {\nu}_{3}^{\prime})^T$,  
defined by 
\be
\nu_f = U^{\prime} \nu^{\prime}, 
\label{basis}
\ee 
where 
\be 
U^{\prime} \equiv U_{23} \Gamma_\delta U_{13}. 
\label{uprime}
\ee
In this basis the Hamiltonian becomes 
\be
{H}^{\prime} =  \frac{1}{2E} U_{12} M^2
U^\dagger_{12} +  U^\dagger_{13} \hat{V} U^\dagger_{13}.
\label{hampr}
\ee 
Taking into account that $V s_{13} c_{13}  \ll  \Delta m_{31}^2/2E$
we can perform the block-diagonalization of ${H}^{\prime}$ 
in (\ref{hampr})  which leads to  
\be
{H}^{\prime} \approx \frac{1}{2E} U_{12} M^2
U^\dagger_{12} +  {\rm diag}(V c_{13}^2, 0, V s_{13}^2).
\label{hampr1}
\ee 
We use notations $c_{13} \equiv \cos \theta_{13}$, $c_{12} 
\equiv \cos \theta_{12}$, 
$s_{12} \equiv \sin \theta_{12}$, 
etc..  Very small term 
$-V c_{13}^2 s_{13}^2 (2VE/\Delta m_{31}^2)$ 
in the 11-element is neglected. 
(In this case the diagonalization is reduced to just 
omitting  the off-diagonal elements in the second term of 
eq. (\ref{hampr}).)  
Essentially, this approximation corresponds 
to neglecting the matter effect on 1-3 mixing.  
The matter correction to the 1-3 mixing equals 
$\theta_{13} (2VE/\Delta m_{31}^2)$, and indeed, 
can be neglected~\cite{Peres:2003wd}. 
According to (\ref{hampr1}), the state ${\nu}_3$  decouples from the
rest of the system and evolves independently. 
The two other states form usual $2\nu-$ system with 
the Hamiltonian $H_2 \equiv H_2(\Delta m_{21}^2, \theta_{12}, V 
c_{13}^2)$.  Therefore the S-matrix (the matrix of  
amplitudes) in the basis $\nu^{\prime}$ has the following form :
\be  
S^{\prime} \approx 
\left(\begin{array}{ccc} 
A_{ee}^{\prime}   & A_{e\mu}^{\prime}    & 0 \\
A_{e\mu}^{\prime}   & A_{\mu\mu}^{\prime}    &   0    \\
0        & 0         & A_{33} 
\end{array}
\right) ~,~~
\label{matr2}
\ee
where 
\be 
A_{33} = \exp(-i 2\phi_{3}^m)\,, \quad \quad      
2 \phi_{3}^m = \frac{\Delta m_{31}^2 L_m}{2E} + V s_{13}^2 L_m,  
\label{phase}
\ee
and  $L_m$ is the total distance traveled by neutrinos in matter. 
The amplitude $A_{e\mu}^{\prime}$ describes transition 
$\nu_\mu^{\prime} \rightarrow \nu_e^{\prime}$. 
The off-diagonal elements of $S^{\prime}$ equal each other:  
\be
A_{\mu e}^{\prime} = A_{e\mu}^{\prime}, 
\ee 
as a consequence of the T-symmetry (the CP-conservation in this basis and 
symmetry of the Earth matter profile).  
The amplitudes $A_{ee}^{\prime}$, $A_{e\mu}^{\prime}$ 
and $A_{\mu\mu}^{\prime}$ are solutions of the evolution equation 
with the Hamiltonian $H_2$. 
As a consequence of unitarity,   
\be
A_{\mu\mu}^{\prime} = A_{ee}^{\prime *}, ~~~~
A_{e\mu}^{\prime *}  = - A_{e\mu}^{\prime}, 
\label{proper}
\ee
or ${\rm Re} A_{e\mu}^{\prime} = 0$,  that is, the transition amplitude 
is pure imaginary. 

According to (\ref{basis}) the $S-$matrix in the original flavor basis 
equals 
\be
S_f = U^{\prime} S^{\prime} U^{\prime \dagger}. 
\ee
Therefore the $\nu_\alpha \rightarrow \nu_\beta$ oscillation probability
is given by 
\be
P(\nu_\alpha \rightarrow \nu_\beta) = 
|(U^{\prime} S^{\prime} U^{\prime \dagger})_{\beta \alpha}|^2. 
\label{abprob}
\ee
Using $U^{\prime}$ defined in (\ref{uprime}) and  
$S^{\prime}$ from (\ref{matr2}), and averaging the oscillations 
related to the third mass eigenstate we obtain explicitly
\begin{eqnarray}
P(\nu_e \rightarrow \nu_e) & = & {c}_{13}^4 |A_{ee}^\prime|^2 + 
{s}_{13}^4,   
\label{ee-prv}\\
P(\nu_\mu \rightarrow \nu_e) & = & {c}_{13}^2\left| 
-{s}_{13} s_{23} e^{- i\delta} A_{ee}^\prime  + 
c_{23} A_{e \mu}^\prime 
\right|^2 + {s}_{13}^2 {c}_{13}^2 s_{23}^2,   
\label{mue-pr}
\end{eqnarray}
and for the inverse channel: 
$P(\nu_e \rightarrow \nu_\mu) =  P(\nu_\mu \rightarrow \nu_e)(\delta 
\rightarrow -\delta)$. Finally, 
\be
P(\nu_\mu \rightarrow \nu_\mu) = \left| 
c_{23}^2 A_{\mu \mu}^\prime
-{s}_{13} \cos \delta \sin 2\theta_{23} A_{e\mu}^\prime + 
s_{13}^2 s_{23}^2 A_{ee}^\prime \right|^2 
+ {c}_{13}^4 s_{23}^4.  
\label{mumu-pr}
\ee
Let us introduce three functions 
\be
{\bf D} \equiv \{P_2, R_2, I_2\}
\ee
(essentially the elements of the density matrix) as
\be
P_2 \equiv |{A}_{\mu e}^\prime|^2  =  1 - |{A}_{ee}^\prime|^2,~~~ 
R_2 \equiv {\rm Re}({A}_{e \mu }^{\prime *}{A}_{ee}^\prime), ~~~~
I_2 \equiv {\rm Im}({A}_{e \mu }^{\prime *} {A}_{ee}^\prime). 
\label{p-r-i}
\ee
These functions satisfy the relation  
\be
P_2^2 + R_2^2 + I_2^2 = P
\ee
which follows from unitarity of the $S-$ matrix. 
The other properties of ${\bf D}-$functions have been studied in 
\cite{Peres:2003wd}. 

According to  (\ref{proper}) and (\ref{p-r-i})  
\be
{\rm Re}({A}_{e \mu }^{\prime *}{A}_{\mu \mu}^\prime) = - R_2, ~~~
{\rm Re}({A}_{\mu \mu }^{\prime *}{A}_{ee}^\prime) = \frac{1}{P_2} (I_2^2 
- R_2^2). 
\label{e-mumu}
\ee
Using these equalities and notations (\ref{p-r-i})  
the probabilities (\ref{ee-prv}), (\ref{mue-pr}) 
and (\ref{mumu-pr}) can be rewritten in terms of ${\bf D}$ as
\beq
P(\nu_e \rightarrow \nu_e) & = & {c}_{13}^4 (1 - P_2) + {s}_{13}^4,    
\label{ee-pr2} \\
P(\nu_\mu \rightarrow \nu_e) & = &
{c}_{13}^2 c_{23}^2 P_2  
- {s}_{13} {c}_{13}^2 \sin 2\theta_{23} (\cos \delta  R_2 + 
\sin \delta I_2) 
\nonumber\\
& + & {s}_{13}^2 {c}_{13}^2 s_{23}^2 (2 - P_{2}), 
\label{mue-pr2}
\eeq  
(which coincides up to the sign of $\delta$ with expression in the  
paper \cite{Peres:2003wd})  
\footnote{In our previous paper \cite{Peres:2003wd} we used different definition of 
amplitudes, 
$A_{\mu e}$ for transition $\nu_\mu \rightarrow \nu_e$, instead of  $A_{e \mu}$.}, 
and  
\beq
P(\nu_\mu \rightarrow \nu_\mu) & =  & 1 - 0.5\sin^2 2\theta_{23}  - c_{23}^4 P_2 +  
2 s_{13}\cos \delta \sin 2\theta_{23} (c_{23}^2 - s_{13}^2 s_{23}^2)R_2 +
\nonumber\\
& + & s_{13}^2 \sin^2 2\theta_{23}
\left[ P_2 \cos^2 \delta  + \frac{1}{2P_2} (I^2_2 - R_2^2)
\right] - 2 s_{13}^2 s_{23}^4 +   
\nonumber\\
& + & s_{13}^4 s_{23}^4 (2 - P_2).  
\label{mumu-pr3}
\eeq
The first line in this equation is a sum of the averaged standard  
$2\nu$-oscillation probability $\nu_\mu \rightarrow \nu_\mu$ 
 driven by the 2-3 mixing, the contribution from the 1-2 oscillations for 
$s_{13} = 0$ 
and the first (linear in $s_{13}$) correction due to the 1-3 mixing  
(the ``induced'' interference \cite{Peres:2003wd}).    
The probability (\ref{mumu-pr3}) is an even function of $\delta$. Notice that in the 
second 
order the correlation of $s_{13}$ and $\delta$:  $s_{13}\cos \delta$ is 
broken and these two parameters 
enter differently, which, in principle,  opens  a possibility to 
determine them independently. 

For antineutrinos we have the same expressions for the probabilities 
with substitutions: $\delta \rightarrow - \delta$, 
$P_2 \rightarrow  \bar{P}_2$, $R_2 \rightarrow  \bar{R}_2$, 
$I_2 \rightarrow  \bar{I}_2$, where 
$\bar{P}_2  \equiv P_2 (V \rightarrow - V)$, etc..

Some insight into properties of the probabilities and their 
dependence on the oscillation parameters  can be obtained  
using  expressions in the constant density case: 
\be
P_2^c = \sin^2 2\theta^m_{12} \sin^2 \phi^m, ~~~
R_2^c = -\frac{1}{2} \sin 4 \theta^m_{12} \sin^2 \phi^m, ~~~ 
I_2^c = \frac{1}{2} \sin 2\theta^m_{12} \sin 2\phi^m. 
\label{prob-con}
\ee
where subscript ``c'' refers to the case of constant density,  
and $\phi^m$ is the half-phase of neutrino oscillations in matter 
defined in (\ref{phase-m}). 
The probabilities for antineutrinos, $\bar{P}_2^c$, $\bar{R}_2^c$ and 
$\bar{I}_2^c$,  can be obtained from  those in eq. (\ref{prob-con}) 
substituting $\theta_m \rightarrow \bar{\theta}_m$ and 
$\phi^m \rightarrow \bar{\phi}^m$.  
The formulas for constant density (\ref{prob-con}) 
give good qualitative and in many cases quantitative  description of the results. 

We compute the $2\nu-$ probabilities $P_2$, $R_2$ and  $I_2$ 
and similar probabilities for antineutrinos  numerically.  
A very precise semi-analytical description of 
the probabilities can be obtained with the adiabatic Magnus expansion. 
In the first order of this expansion using the results of \cite{ara9} 
(see eq. (78) and (76) in \cite{ara9}) we obtain for the amplitudes in the 
$\nu^\prime$ basis 
\beq
A_{ee}^\prime & = & \cos I_\theta \cos \phi^{ad} + 
i \left(\cos I_\theta \sin \phi^{ad} \cos 2\theta_{12}^0  
- \sin I_\theta \sin 2 \theta_{12}^0 \right), 
\label{ad-ee}\\
A_{e\mu}^\prime & = &  -i \left(\sin I_\theta \cos 2 \theta_{12}^0 
+ \cos I_\theta ~\sin\phi^{ad} \sin 2\theta_{12}^0 \right),   
\label{ad-emu}
\eeq
where $\theta_{12}^0$ is the mixing angle in matter at the 
surface of the Earth, $\theta_{12}^0 \equiv \theta^m (x_f)$, 
$x_f$ is the surface coordinate, and    
\be 
I_\theta = - 2 \int^{x_f}_{\bar{x}} dx \left[\frac{d \theta^m_{12}(x)}{dx}\right] 
\sin \phi^{ad} (x).  
\ee 
Here $\bar{x}$ is the central point of the trajectory,
and the expressions for probabilities are valid for a symmetric 
(with respect to $\bar{x}$)  density profile.  The adiabatic half-phase 
is given by  
\be
\phi^{ad}(x) \equiv  \frac{\Delta m^2_{21}}{4E}
\int^{x}_{\bar{x}} dr \sqrt{\left(\cos 2\theta_{12} - 
\frac{2E V(r) c_{13}^2}{\Delta m^2_{21}} \right)^2 + 
\sin^2 2\theta_{12}}  
\ee
and in eq.(\ref{ad-ee}, \ref{ad-emu}) $\phi^{ad} \equiv \phi^{ad}(x_f)$.   

Using the amplitudes (\ref{ad-ee}, \ref{ad-emu}) we obtain the following 
expressions for $P_2, R_2, I_2$:   
\beq 
P_2 & = & \cos^2 I_\theta \sin^2 2\theta_{12}^0 \sin^2 \phi^{ad} 
+ \frac{1}{2} \sin 2 I_\theta \sin 4\theta_{12}^0 \sin \phi^{ad} 
+ \sin^2 I_\theta  \cos^2 2 \theta_{12}^0, 
\nonumber\\
R_2 & = &- \frac{1}{2} \left[\cos^2 I_\theta \sin 4\theta_{12}^0 \sin^2 \phi^{ad} 
+ \sin 2 I_\theta \cos 4\theta_{12}^0 \sin \phi^{ad}  -
\sin^2 I_\theta \sin 4\theta_{12}^0 \right], 
\nonumber\\
I_2 & = & \frac{1}{2} \left[\cos^2 I_\theta \sin 2\theta_{12}^0 \sin 2\phi^{ad} 
+ \sin 2 I_\theta \cos \phi^{ad} \cos 2\theta_{12}^0 \right].  
\label{pri-ad}
\eeq
These expressions are  reduced to the expressions   
in (\ref{prob-con}) with substitutions
$\theta_{12}^m \rightarrow \theta_{12}^0$ and  
$\phi^{ad} \rightarrow  \phi^m$ if $I_\theta = 0$ which would correspond to 
a perfect adiabaticity. From (\ref{pri-ad})  we find formulas with the first 
non-adiabatic corrections 
\begin{eqnarray}
P_2 & \approx &  \sin^2 2\theta_{12}^0 \sin^2 \phi^{ad} + 
I_\theta \sin 4\theta_{12}^0 \sin \phi^{ad},  
\nonumber\\ 
R_2 & \approx & - \frac{1}{2} \sin 4\theta_{12}^0 \sin^2 \phi^{ad} 
- I_\theta \cos 4\theta_{12}^0 \sin \phi^{ad},  
\nonumber \\
I_2 & \approx & \frac{1}{2} \sin 2\theta_{12}^0 \sin 2\phi^{ad} 
+ I_\theta \cos 2\theta_{12}^0 \cos \phi^{ad}. 
\end{eqnarray}
For completeness in the Appendix A we present also results in the second 
order of the Magnus expansion which provides better than $1\%$ accuracy 
for all neutrino energies. 

The results of this section can be used immediately for high energy neutrinos 
$E > 150$ MeV and for trajectories far from horizon, when oscillations 
in the atmosphere can be neglected. 

\section{Oscillation effects in the atmosphere and inside the Earth}

Above horizon the oscillations occur in vacuum (we neglect density of the 
atmosphere).  The evolution of neutrinos can be considered again  
in the $\nu^\prime$ basis, since the connecting matrix 
$U^{\prime}$ depends on the vacuum mixing angles only.

Again, in the $\nu^\prime$ basis
the third mass state decouples and for two other states 
the $2\times 2$ block of the S-matrix is given by  
\be  
S_A^{(2)} \approx 
\left(\begin{array}{ccc} 
c_{\phi} + i \cos 2\theta_{12} s_\phi  & - i \sin 2\theta_{12} s_\phi    \\
- i \sin 2\theta_{12} s_\phi   &    c_{\phi} - i \cos 2\theta_{12} s_\phi 
\end{array}
\right) ~.~~
\label{matrA}
\ee
Here $s_\phi \equiv \sin \phi$, $c_\phi \equiv \cos \phi$ and the half-phase 
$\phi$ is defined in (\ref{phase-m}). The oscillation probabilities in the   
flavor basis are given by the  expressions (\ref{ee-pr2} - \ref{mumu-pr3})
with $P_2 = \sin^2 2\theta_{12} \sin^2 \phi$, 
$R_2 = -\frac{1}{2} \sin 4 \theta_{12} \sin^2 \phi$, and  
$I_2 = \frac{1}{2} \sin 2\theta_{12} \sin 2\phi$.

For trajectories below the horizon one needs to take into account oscillations   
in two layers: the atmosphere and  the Earth. 
For these trajectories  
the total $S-$ matrix in the $\nu^\prime-$basis equals
\be  
S_{tot} = S^\prime \cdot S_A  \approx 
\left(\begin{array}{ccc} 
\tilde{A}_{ee}   & \tilde{A}_{e\mu}     & 0 \\
\tilde{A}_{\mu e}   & \tilde{A}_{\mu\mu}    &   0    \\
0        & 0         & \tilde{A}_{33} 
\end{array}
\right) ~,~~
\label{s-tot}
\ee
where $S^\prime$ is given in (\ref{matr2}) 
and $S_A$ describes evolution in atmosphere. 

Using the expressions for the amplitudes (\ref{tilde-a}) from the Appendix B  
and properties of the amplitudes for a single symmetric layer 
(\ref{proper}) we obtain the relations
\be
\tilde{A}_{\mu\mu} =  \tilde{A}_{ee}^*, ~~~~
\tilde{A}_{\mu e} = - \tilde{A}_{e\mu}^* .
\label{proper2}
\ee

The flavor oscillation probabilities can be computed according to  
$P(\nu_\alpha \rightarrow \nu_\beta) = 
|(U^{\prime} S_{tot} U^{\prime \dagger})_{\beta \alpha}|^2$.    
As a result, we obtain formulas  similar to those in 
(\ref{ee-pr2} - \ref{mumu-pr3}) for a single layer with, 
essentially, substitution 
$A_{\alpha \beta}^{\prime} \rightarrow  \tilde{A}_{\alpha \beta}$:  
\beq
P(\nu_e \rightarrow \nu_e) & = & {c}_{13}^4 |\tilde{A}_{ee}|^2 + {s}_{13}^4,   
\label{ee-pr}\\
P(\nu_\mu \rightarrow \nu_e) & = & {c}_{13}^2\left| 
-{s}_{13} s_{23} e^{- i\delta} \tilde{A}_{ee}  + 
c_{23} \tilde{A}_{e \mu} \right|^2 + {s}_{13}^2 {c}_{13}^2 s_{23}^2,\\   
P(\nu_e \rightarrow \nu_\mu) & = & {c}_{13}^2\left| 
-{s}_{13} s_{23}  e^{i\delta} \tilde{A}_{ee} + 
c_{23} \tilde{A}_{\mu e } \right|^2 + {s}_{13}^2 {c}_{13}^2 s_{23}^2,\\   
P(\nu_\mu \rightarrow \nu_\mu) & = & \left| c_{23}^2 \tilde{A}_{\mu \mu}
- i {s}_{13} \sin 2\theta_{23}{\rm Im}(e^{ i\delta} \tilde{A}_{e\mu}) +  
s_{13}^2 s_{23}^2 \tilde{A}_{ee}\right|^2 
+ {c}_{13}^4 s_{23}^4.  
\label{4prob}
\eeq
In the last equation  we used the properties (\ref{proper2}).  
 
Since the two-layer density profile is not symmetric, we have  
inequality $\tilde{A}_{\mu e} \neq \tilde{A}_{e\mu}$,  and 
furthermore,  these amplitudes are not pure imaginary. 
Consequently, the probabilities 
(\ref{ee-pr2} -\ref{4prob}) are expressed now in terms of 5 different functions 
\be
\tilde{\bf D} \equiv  \{ \tilde{P}, R_{e\mu}, R_{\mu e}, I_{e\mu}, I_{\mu e} \}, 
\ee 
and not three,  $\bf D$, as in the case of symmetric 
profile. We introduce them as 
\beq
\tilde{P}~~ & \equiv & |\tilde{A}_{\mu e}|^2 = |\tilde{A}_{e \mu}|^2, ~~~  
R_{e\mu}  \equiv  {\rm Re}(\tilde{A}_{e \mu }^* \tilde{A}_{ee}),~~~
I_{e\mu} \equiv {\rm Im}(\tilde{A}_{e \mu }^* \tilde{A}_{ee}), 
\nonumber\\
R_{\mu e} & \equiv & {\rm Re}(\tilde{A}_{\mu e}^* \tilde{A}_{ee}),~~~
I_{\mu e}  \equiv  {\rm Im}(\tilde{A}_{\mu e}^* \tilde{A}_{ee}). 
\label{5funct}
\eeq
From unitarity of the $S-$ matrix we have 
$|\tilde{A}_{\mu \mu}|^2 = |\tilde{A}_{ee}|^2 = 1 - \tilde{P}$
and $\tilde{A}_{e e} \tilde{A}_{\mu e}^* = 
- \tilde{A}_{e \mu} \tilde{A}_{\mu \mu}^*$. 
In turn, $\tilde{P}$, $R_{e\mu}$,  $I_{e\mu}$,  $R_{\mu e}$ and  
$I_{\mu e}$ are the functions of $P_2$, $R_2$, $I_2$ and vacuum 
the oscillation parameters $\phi$ and $\theta_{12}$. We present these expressions 
in the Appendix B. 

In terms of $\tilde{\bf D}-$functions  (\ref{5funct}) the expressions 
for probabilities are  similar to those for 1 layer 
(\ref{ee-pr2} - \ref{mumu-pr3}) with, however, certain  
differences.  
Using (\ref{ee-pr} - \ref{4prob}) and definitions (\ref{5funct}) 
we obtain 
\be
P(\nu_e \rightarrow \nu_e) =  c_{13}^4 (1 - \tilde{P})  + s_{13}^4 ,    
\label{ee-pr3t}
\ee
\be
P(\nu_\mu \rightarrow \nu_e) = 
{c}_{13}^2 c_{23}^2 \tilde{P}  
- {s}_{13} {c}_{13}^2 \sin 2\theta_{23} (\cos \delta  R_{e\mu} + 
\sin \delta I_{e \mu}) 
+ {s}_{13}^2 {c}_{13}^2 s_{23}^2 (2 - \tilde{P}). 
\label{mue-pr2t}
\ee  
Then 
\be
P(\nu_e \rightarrow \nu_\mu) = 
{c}_{13}^2 c_{23}^2 \tilde{P}  
- {s}_{13} {c}_{13}^2 \sin 2\theta_{23} (\cos \delta  R_{\mu e} -  
\sin \delta I_{\mu e}) 
+ {s}_{13}^2 {c}_{13}^2 s_{23}^2 (2 - \tilde{P}).  
\label{emu-pr2t}
\ee 
The most significant  change is in the $\nu_\mu \rightarrow \nu_\mu$ probability:  
\beq
P(\nu_\mu \rightarrow \nu_\mu) & = &  1 - 0.5 \sin^2 2\theta_{23} 
- c_{23}^4 \tilde{P} +   
\nonumber\\
& + & s_{13} \sin 2\theta_{23} (c_{23}^2 - s_{13}^2 s_{23}^2)  
[\cos \delta(R_{e\mu} + R_{\mu e}) + \sin \delta (I_{e\mu} - I_{\mu e})] + 
\nonumber\\ 
\, & + & s_{13}^2 \sin^2 2\theta_{23} \frac{\tilde{P}}{2} 
\left\{ 1 + \frac{1}{\tilde{P}(1 - \tilde{P})}
\left[\cos 2\delta (R_{e\mu} R_{\mu e} + I_{e\mu}I_{\mu e}) +  \right. \right.
\nonumber \\
& + & \left. \left. \sin 2\delta (R_{\mu e} I_{e \mu} - R_{e\mu}I_{\mu e})
\right]
  +  \frac{1}{\tilde{P}^2} (I_{e\mu} I_{\mu e} - R_{e\mu}R_{\mu e})
\right\} - 
\nonumber\\
  & -  & 2 s_{13}^2 s_{23}^4  +  s_{13}^4 s_{23}^4 (2 - \tilde{P}).     
\label{mumu-til}
\eeq
In the limit $\phi \rightarrow 0$ (no oscillations 
in atmosphere) we have
\be 
R_{e\mu} = R_{\mu e} = R_2,~~ I_{e\mu} = I_{\mu e} = I_2, ~~ 
\tilde{P} = P_2,  
\ee
and the formulas (\ref{ee-pr3t} - \ref{mumu-til}) are reduced 
to the one-layer formulas of the previous section. 
Notice that functional dependence of the probabilities on the parameters 
$s_{13}$, $\delta$ and $\theta_{23}$ is practically the same as 
in the one layer case. 

Let us consider expressions for  
$R_{e\mu}$, $R_{\mu e}$,  $I_{e\mu}$, $I_{\mu e}$ and 
$\tilde{P}$ in the constant density approximation   
(see Appendix C.) 
They can be presented in the following form 
\be 
\tilde{\bf D}_i^c  = {\bf D}_i^c (\phi^m + \phi) +  
\sin \Delta \theta ~\chi_i(\theta_{12}^m, \theta_{12}), 
\label{form+}
\ee 
where $\Delta \theta \equiv \theta_{12}^m - \theta_{12}$. 
So, $\tilde{\bf D}_i^c$ can be written as 
the corresponding functions for one layer of matter with total phase 
$\phi^m + \phi$ plus corrections related to difference 
of mixing angles in matter and vacuum. 
Since at low energies $\Delta \theta \ll  \theta_{12}$,   
the second term  in (\ref{form+}) can be considered as small correction. 
In the first order in $\Delta \theta$ the expressions (\ref{pri-const}) 
become  
\beq
\tilde{P}^c~ & \approx & \sin^2 2\theta_{12}^m \sin^2 (\phi^m + \phi) 
 - 2 \Delta \theta  \sin 4 \theta_{12}  \sin (\phi^m + \phi)   
 \sin \phi \cos \phi^m , 
\nonumber\\
R_{e \mu}^c & \approx &
- \frac{1}{2} \sin 4\theta^m_{12} \sin^2 (\phi^m + \phi)
+ \Delta \theta \left[- \sin^2 2\theta^m_{12} \sin 2\phi \sin 2\phi^m + \right. 
\nonumber\\
& + & \left. 2 \sin^2\phi (\cos 4\theta_{12} \cos^2 \phi^m +  
\sin^2 \phi^m) \right] ,  
\nonumber\\
R_{\mu e }^c & \approx &
- \frac{1}{2} \sin 4\theta^m_{12} \sin^2 (\phi^m + \phi)
+ \Delta \theta \left[\cos^2 2\theta^m_{12} \sin 2\phi \sin 2\phi^m + \right.
\nonumber\\
& + & \left. 2 \sin^2\phi (\cos 4\theta_{12} \cos^2 \phi^m  - 
\sin^2 \phi^m) \right] ,
\nonumber\\
I_{e \mu}^c & \approx & 
\frac{1}{2} \sin 2\theta^m_{12}  \sin 2(\phi^m + \phi)
- \Delta \theta \cos 2 \theta^m_{12} \sin 2\phi, 
\nonumber\\
I_{\mu e}^c & \approx &
\frac{1}{2} \sin 2\theta^m_{12}  \sin 2(\phi^m + \phi) + 
\nonumber\\
& + &\Delta \theta \cos 2\theta^m_{12} 
\left[-\sin 2\phi \cos 2\phi^m + 2 \sin^2 \phi \sin 2\phi^m \right]. 
\eeq
At energies $E < 60$ MeV the magnitude of second term does not exceed 0.1. 
For high energies the difference $\Delta \theta$ becomes large. However in this case 
the oscillation length increases and  $\phi$ becomes smaller, so that 
the oscillations in atmosphere can be neglected. 
 
Summarizing, final  expressions for the flavor probabilities 
are given in eqs. (\ref{ee-pr3t}), (\ref{mue-pr2t}),(\ref{emu-pr2t})
(\ref{mumu-til}) with  $\tilde{P}$, $R_{e\mu}$, $I_{e\mu}$, 
$R_{\mu e}$, $I_{\mu e}$ as functions of $P_2$, $R_2$ and $I_2$, 
and the vacuum oscillation parameters presented in the Appendix B. 
In turn, the precise semi-analytical expressions for $P_2$, $R_2$ and 
$I_2$ are given in (\ref{pri-ad}) for the first order Magnus expansion  and 
in the Appendix A -- for the second order.

\section{Averaging of oscillations and integration over the zenith angle}

Results for the oscillation probabilities obtained in the previous 
section are  simplified substantially after integration over 
the zenith angle. In what follows we will consider  detection of 
neutrinos by the charged current interactions with nucleons, where the 
information about neutrino 
direction is essentially lost. Consequently,  observables  
(at least in the first approximation) are given by integration over 
the zenith angle. To a good approximation the neutrino flux at low 
energies does not depend on the zenith angle 
$\Theta_\nu$,  and therefore  integration over the angle 
is reduced to averaging of probabilities over  $\Theta_\nu$.  

In the constant density approximation 
the probabilities averaged over the oscillation phase  
equal 
\be
\langle P_2^c \rangle_{\phi} = \frac{1}{2} \sin^2 2\theta_{12}^m~, ~~~
\langle R_2^c \rangle_{\phi} = -\frac{1}{4} \sin 4 \theta_{12}^m, ~~~ 
\langle I_2^c \rangle_{\phi} =  0. 
\label{avPI}
\ee
where subscript means averaging over the phase. 
For antineutrinos one should substitute 
$\theta_{12}^m  \rightarrow \bar{\theta}_{12}^m$. 
As a consequence of  $I_2^c = 0$ the CP-odd terms of probabilities 
(which lead to the CP-asymmetries) become zero, as is expected,  
since this would correspond to averaging over all 
oscillation phases (we have already averaged over the 1-3 phase). 
The real part $R_2$ determines 
the first order correction due to the 1-3 mixing.

For very low energies we have 
$\theta_{12}^m \approx \theta_{12}$,  and consequently, 
$\langle R_2 \rangle_{\phi} = \langle \bar{R}_2 \rangle_{\phi} =
-\frac{1}{4} \sin 4 \theta_{12} \approx - 0.19$. 
For antineutrinos with increase of energy 
the angle $\bar{\theta}_{12}^m$ decreases. 
As a result, $\langle \bar{R}_2 \rangle_\phi < 0$ in whole energy range. 
The absolute value 
$|\langle \bar{R}_2 \rangle_{\phi}|$ first increases, reaches maximal   
value 0.25 when $\bar{\theta}_{12}^m = 22.5^{\circ}$ 
(at $E \sim 150$ MeV) and then decreases.  In contrast, 
the angle $\theta_{12}^m$
increases, and therefore $\langle R_2 \rangle_{\phi}$ first decreases, 
becomes zero in the resonance ($E \sim 100$ MeV) and then  changes the 
sign: $\langle R_2 \rangle > 0$. It increases with energy till 
$E \sim 300$ MeV  and then decreases again.  

For the two layer case we obtain after averaging over the phases  
\beq
\langle \tilde{P}^c \rangle_{\phi}~~ & = & \frac{1}{2}\sin^2 2\theta^m_{12}
- \frac{1}{4} (\sin^2 2\theta_{12}^m - \sin^2 2\theta_{12}) + 
 +  \frac{1}{4}\sin^2 2(\theta_{12} - \theta^m_{12}), 
\nonumber\\
\langle R_{e \mu}^c \rangle_{\phi} & = & - \frac{1}{4} \sin 4\theta^m_{12} 
+ \frac{1}{8} \left[\sin 4\theta^m_{12} - \sin 4\theta_{12}
+ \sin 4(\theta^m_{12} -\theta_{12})\right], 
\nonumber\\
\langle R_{\mu e}^c \rangle_{\phi} & = & - \frac{1}{4} \sin 4\theta^m_{12}
+ \frac{1}{8} \left[\sin 4\theta^m_{12} - \sin 4\theta_{12}
- \sin 4(\theta^m_{12} -\theta_{12})\right],  
\nonumber\\
\langle I_{e \mu}^c \rangle_{\phi} & = & \langle I_{\mu e}^c \rangle_{\phi} = 0.  
\label{tildew}
\eeq

At very low energies, when the difference of mixing angles 
in vacuum and in matter is very small, we find from (\ref{tildew}):  
\beq
\langle \tilde{P} \rangle_{\phi} ~~& \approx &  \frac{1}{2}\sin^2 
2\theta^m_{12} - \frac{1}{2} \Delta \theta \sin 4\theta_{12}, 
\nonumber \\
\langle R_{e \mu}\rangle_{\phi} & = & 
- \frac{1}{2} \sin 4\theta^m_{12} + \Delta\theta \cos^2 2\theta_{12} , 
\nonumber\\
\langle R_{\mu e} \rangle_{\phi} & = &
- \frac{1}{2} \sin 4\theta^m_{12} - \Delta\theta \sin^2 2\theta_{12}. 
\label{avcorr}
\eeq 

These averaged functions  $\langle D_i \rangle_{\phi}$ determine the number 
of low energy events. Indeed, 
to find the number of events we need to integrate 
the probabilities folded with the neutrino fluxes  and cross-sections 
over the zenith angle $\Theta_\nu$ 
(the angular variables,  in general). In turn, 
the oscillation probabilities are linear functions of $\tilde{\bf D}_i$.  
Therefore we deal here with integrals  
\be
J_i = \int d(\cos \Theta_\nu) \tilde{D}_i(\cos\Theta_\nu) F(\Theta_\nu). 
\ee
At low energies the fluxes do not depend on $\Theta_\nu$, 
and consequently,  
\be
J_i = 2F ~\langle \tilde{D}_i(\cos\Theta_\nu) \rangle.
\ee
Here  $\langle ... \rangle$ without subscript denotes averaging over 
the zenith angle $\Theta_\nu$. The functions 
$D_i$ depend on $\cos\Theta_\nu$ via the 
oscillation phases $\phi$ and $\phi_m$. 
Apparently $\phi \propto L_A (\cos\Theta_\nu)$, 
and in the constant density approximation,  
$\phi_m \propto L_m \propto \cos\Theta_\nu$. 
Therefore averaging over $\cos\Theta_\nu$
is equivalent to averaging over  the phase $\phi_m$ as we did 
above. The situation is different for the terms which 
depend on the vacuum phase. Indeed, according to 
(\ref{totlen}) the dependence of $\phi$ on $\cos\Theta_\nu$  
is non-linear. In particular, for trajectories not very  
close to the horizon $L_A \propto 1/\cos\Theta_\nu$ 
and therefore integration over $\cos\Theta_\nu$ is not 
reduced to  averaging over the phase. 
For estimations, the effect of oscillations 
in atmosphere can be taken into account by the additional factor $\kappa$ as 
\be
\langle \tilde{\bf D}_i(\cos\Theta_\nu) \rangle  \approx 
\frac{1}{2}\int d(\cos\Theta_\nu) ~\tilde{\bf D}_i  =  
\frac{\kappa}{2} \langle \tilde{\bf D}_i \rangle_{\phi},  
\label{d-func}
\ee
where $\langle D_i \rangle_{\phi}$ refers to averaging over the phase. 
The factor  $\kappa$ equals 1,  if oscillations in  atmosphere can be neglected. 
For low energy bins ($E < 15$ MeV)  $\kappa$ can be as big as 1.2; 
for $E = 20$ MeV we have $\kappa \sim 1.1$ and $\kappa$  approaches 1 with 
increase of energy.

\section{Neutrino fluxes at the detector}

The  $\nu_e-$flux at the detector 
with oscillations taken into account, $F_e$,  
can be written as 
\be
F_e = F_e^0 P(\nu_e \rightarrow \nu_e) +  F_{\mu}^0 
P(\nu_\mu \rightarrow \nu_e) = 
F_e^0 \left[ P(\nu_e \rightarrow \nu_e) + 
r P(\nu_\mu \rightarrow \nu_e) \right],    
\label{fluxe-osc}
\ee
where $r$ is defined in (\ref{f-ratio}). For antineutrinos we have similar 
expression with substitutions $P \rightarrow \bar{P}$, 
$F \rightarrow \bar{F}$, $r \rightarrow \bar{r}$. 

Inserting the probabilities 
$P_{ee}$ and  $P_{\mu e}$ from (\ref{ee-pr3t}) and 
(\ref{mue-pr2t})  into (\ref{fluxe-osc}) we obtain the  expression 
for the  relative change  of the $\nu_e-$flux due to oscillations:
\beq
\frac{F_e}{F_e^0} & = & 1 + (r c_{23}^2 - 1) \tilde{P} 
- r {s}_{13} {c}_{13}^2 \sin 2\theta_{23} 
(\cos\delta ~ R_{e \mu}  + \sin \delta ~I_{e\mu}) - 
\nonumber\\
& -  & 2 {s}_{13}^2 \left[(1 - r s_{23}^2) + \tilde{P} (r - 2) \right]  
+ {s}_{13}^4 (1 - r s_{23}^2) (2 - \tilde{P}).    
\label{fl-excess}
\eeq
It coincides (up to the sign of $\delta$ and the corresponding change of 
${\bf D} \rightarrow \tilde{\bf D})$   
with the expression in our previous paper~\cite{Peres:2003wd}. 
Notice that the corrections of the order ${s}_{13}^2$  
are suppressed by the ``screening'' factors which are zero for 
maximal 2-3 mixing and $r = 2$. There is no corrections 
of the order $s_{13}^3$.  

The $\nu_\mu-$flux  at the detector with oscillations taken into account, 
$F_{\mu}$, can be written as 
\be
F_{\mu} = F_{\mu}^0  P(\nu_\mu \rightarrow \nu_\mu) +  
F_{e}^0  P(\nu_e \rightarrow \nu_\mu) = 
F_{\mu}^0 \left[ P(\nu_\mu \rightarrow \nu_\mu) + 
\frac{1}{r} P(\nu_e \rightarrow \nu_\mu) \right].     
\label{fluxe-oscd}
\ee
Then the ratio of fluxes  with and without oscillations equals
\beq
\frac{F_\mu}{F_\mu^0} & = & 1 - \frac{1}{2}\sin^2 2\theta_{23} -  
c_{23}^2\tilde{P} \left(c_{23}^2 - \frac{{c}_{13}^2}{r} \right) - 
\nonumber\\
& - & {s}_{13} \sin 2\theta_{23} \left\{\cos \delta 
\left[ \frac{R_{\mu e}}{r} - c_{23}^2 (R_{e\mu} + R_{\mu e}) \right] 
\right. \nonumber\\
& - & \left. \sin \delta  \left[\frac{I_{\mu e}}{r} + 
c_{23}^2 (I_{e\mu} + I_{\mu e})   \right]\right\} + O({s}_{13}^2).
\label{mumu-excess}
\eeq
Averaging over the zenith angle vanishes the imaginary parts.  
Notice that the linear in $s_{13}$ term is proportional to 
$R_{e\mu} + R_{\mu e} \approx  2 R_2$.

\section{Electron neutrino (antineutrino) events}

We will discuss here (mostly for illustration) a  detection of 
the low energy neutrinos in the water Cherenkov detectors.  
In these detectors the electron neutrinos are 
detected  via the  quasi-elastic scattering 
\be
\bar{\nu}_e  +  p \rightarrow e^+ + n, ~~~
\bar{\nu}_e  +   ^{16}O \rightarrow e^+ + N, ~~~
{\nu}_e  +  ^{16}O \rightarrow e^- + F .
\label{barnuo}
\ee

The energy of positron (electron) is practically uniquely related to 
the neutrino energy: 
$E_e = E_\nu - 1.293$ MeV, (for $^1H$), $E_e = E_\nu - 15$ MeV 
(for $^{16}O$). 
The number of $e^+$ events (\ref{barnuo}) has been computed as 
\be
  N_{\bar{e}} \propto \sum_{i = p, O}
 \int  dE_\nu dE_e d(\cos\Theta_\nu)  
\;   F_{\bar e}(E_{\nu},\Theta_{\nu}) \frac{d{ \sigma_i}}{dE_e} 
\Psi(\Theta_e,\Theta_{\nu},E_{\nu})     
\varepsilon(E_e),
\label{event1}
\ee
where $F_{\bar e}$ is the atmospheric $\bar{\nu}_e$-flux
at the detector given in (\ref{fl-excess})  for $\bar{\nu}_e$;  
the fluxes $F_{\bar e }^0$ and $F_{\bar \mu}^0$ without oscillations 
are taken from Refs.~\cite{Gaisser:1988ar,Honda:1995hz,Battistoni:2005pd}; 
$d\sigma_i / dE_{e}$ are the differential 
cross-sections ~\cite{LLS},  
$\varepsilon(E_e)$ is the detection efficiency.  
$\Psi$ is the ``dispersion'' function which describes  
deviation of the lepton zenith angle from the neutrino 
zenith angle~(for details see Ref.~\cite{compute}). 
Summation proceeds over  scattering on Hydrogen and 
Oxygen. For ${\nu}_e$ we use similar expression 
for number of events but consider the reaction on Oxygen only.  

\begin{figure}[htbp]
\includegraphics[scale=0.7]{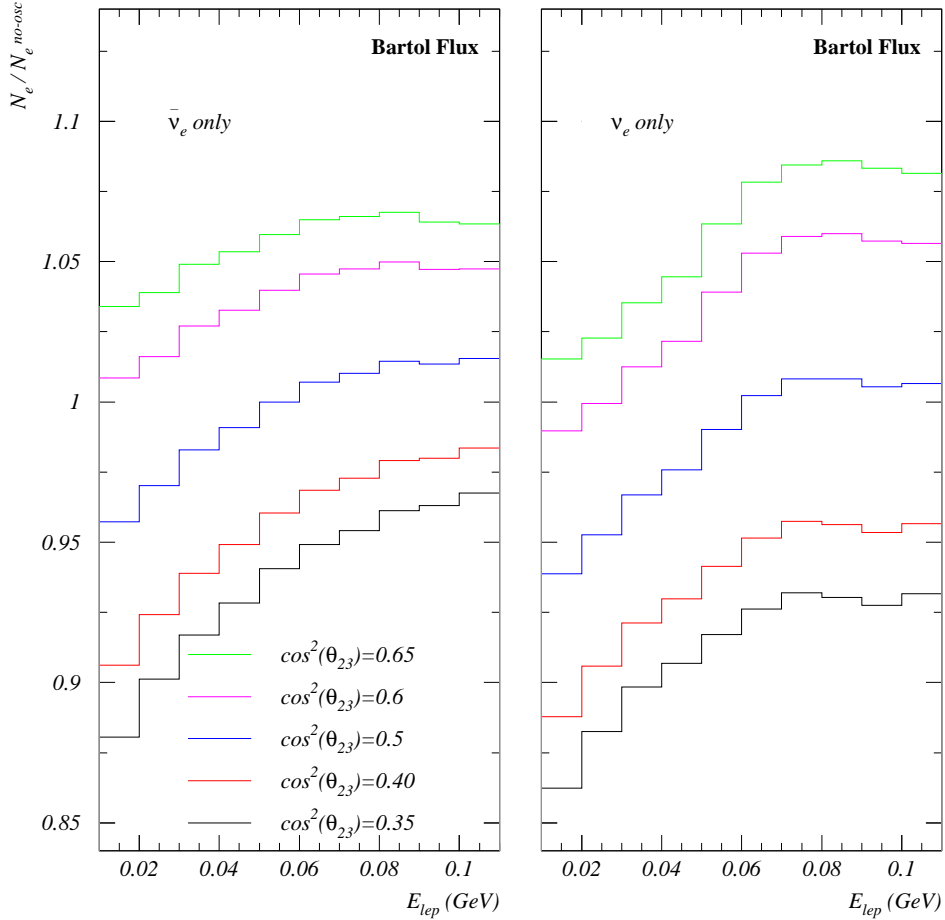}
\caption{ {\it Left panel:} Distortion of the positron  energy spectrum 
due to oscillations of the atmospheric $\bar{\nu}_e$.  
Shown is the ratio $N_{\bar e}/N_{\bar e}^0$ as  function 
of the positron energy for different values of $\cos^2 \theta_{23}$. 
The Bartol fluxes and values of oscillation parameters 
$\sin^2 2\theta_{12}  = 0.82$, $\Delta m^2_{21} = 7.3 \times 10^{-5}$ eV$^2$ 
and  $s_{13} = 0$ used for calculations of the histograms.  
Both interactions with  $^1H$ and $^{16}O$ in a water Cherenkov detector 
are taken into account. {\it Right panel:} The same as in the left panel 
for the atmospheric $\nu_e$ and the electron spectrum. 
The interactions with $^{16}O$ in water Cherenkov 
detectors are taken into account.
}
\label{f4}
\end{figure}

Results of computations of the energy spectra of $\bar{\nu}_e- $ and 
${\nu}_e- $ events for different values of 2-3 mixing  
are shown in fig. \ref{f4}.  
Qualitative and to a large extend quantitative understanding of these results 
can be obtained from the following semi-analytical consideration. 
The number of events can be presented as 
\be
N_e = \left\langle \frac{F_e}{F_e^0} \right\rangle  N_e^0, 
\ee
where $N_e^0$ is the number of events without oscillations; 
the ratio of fluxes $F_e/F_e^0$ is given in 
(\ref{fl-excess}) and  averaging proceeds over the neutrino 
energy bin and the zenith angle.

The relative change of the spectrum can be written in the following way: 
\be
\frac{N_e}{N_e^0} = 1 + \epsilon_e^{(0)} + \epsilon_e^{(1)} + 
\epsilon_e^{(2)} + ... , 
\label{eps}
\ee
where 
\be
\epsilon_e^{(0)} = \langle \tilde{P}(E) \rangle \left[\langle r(E) \rangle  
c_{23}^2 - 1 \right] 
\label{eps0}
\ee
is the correction due to the oscillations driven 
by the 1-2 mixing and split for 
$s_{13} = 0$; 
\be
\epsilon_e^{(1)} = - s_{13} \cos \delta \sin 2\theta_{23}
\langle r \rangle \langle R_{e\mu}\rangle
\label{eps1}
\ee
is the first linear correction due to 1-3 mixing, etc..
Here $ \langle \tilde{P}(E) \rangle $, $ \langle R_{e\mu}(E) \rangle $ 
and  $\langle r(E) \rangle$ are the probabilities   
and the ratio of fluxes averaged 
over the zenith angle and energy bin correspondingly. We take 
$\langle r R_2 \rangle \approx \langle r \rangle \langle R_2 \rangle$.   
For antineutrinos $P_2 \rightarrow \bar{P}_2(E)$ and $r \rightarrow  \bar{r}$.

In fig. \ref{f4} (left panel),  we show  distortion of the energy spectrum 
of  positrons produced by the atmospheric $\bar{\nu}_e$ 
for $s_{13} = 0$. All the features of the curves can be traced 
from  eqs. (\ref{eps}, \ref{eps0}). 
The averaged probability $\langle \bar{P}_2(E) \rangle$ slightly decreases 
with energy. For estimations one can use the oscillation probability 
for a layer of constant density, so that $\langle \bar{P}_2(E) \rangle 
\sim 0.25 \kappa \sin^2 2\theta_m$,  where  $\kappa$ takes into account 
oscillations in atmosphere (\ref{d-func}).

\begin{figure}[htbp]
\includegraphics[scale=0.7]{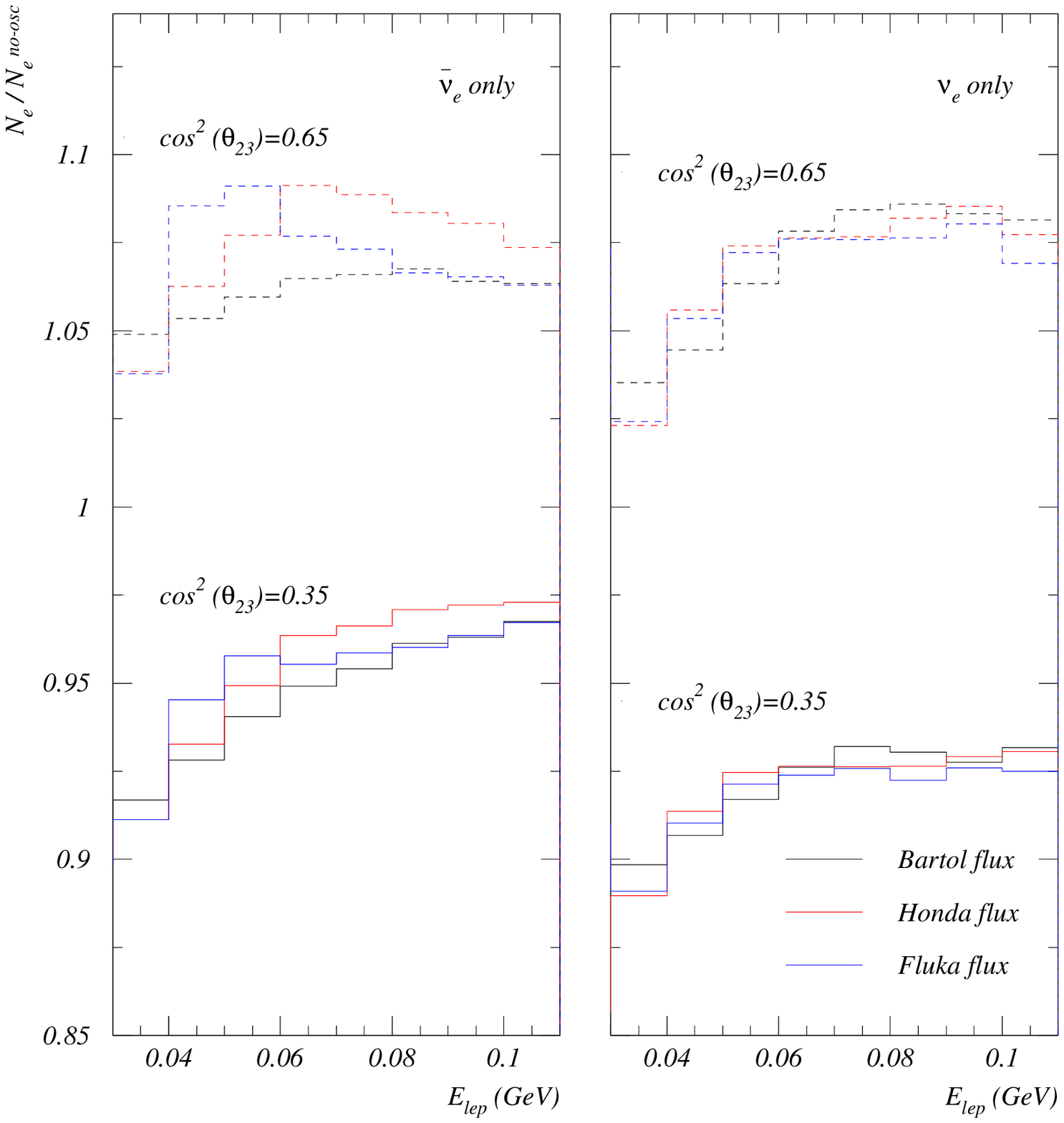}
\caption{Distortions of the energy spectra of events for  
the neutrino fluxes from different computations. 
The full and dashed lines correspond to two different 
values of $\cos^2 \theta_{23}$. {\it Left panel}: $\bar{\nu}_e$, 
{\it Right panel:} ${\nu}_e$. 
}
\label{diff-sp}
\end{figure}

\begin{figure}[htbp]
\includegraphics[scale=0.55]{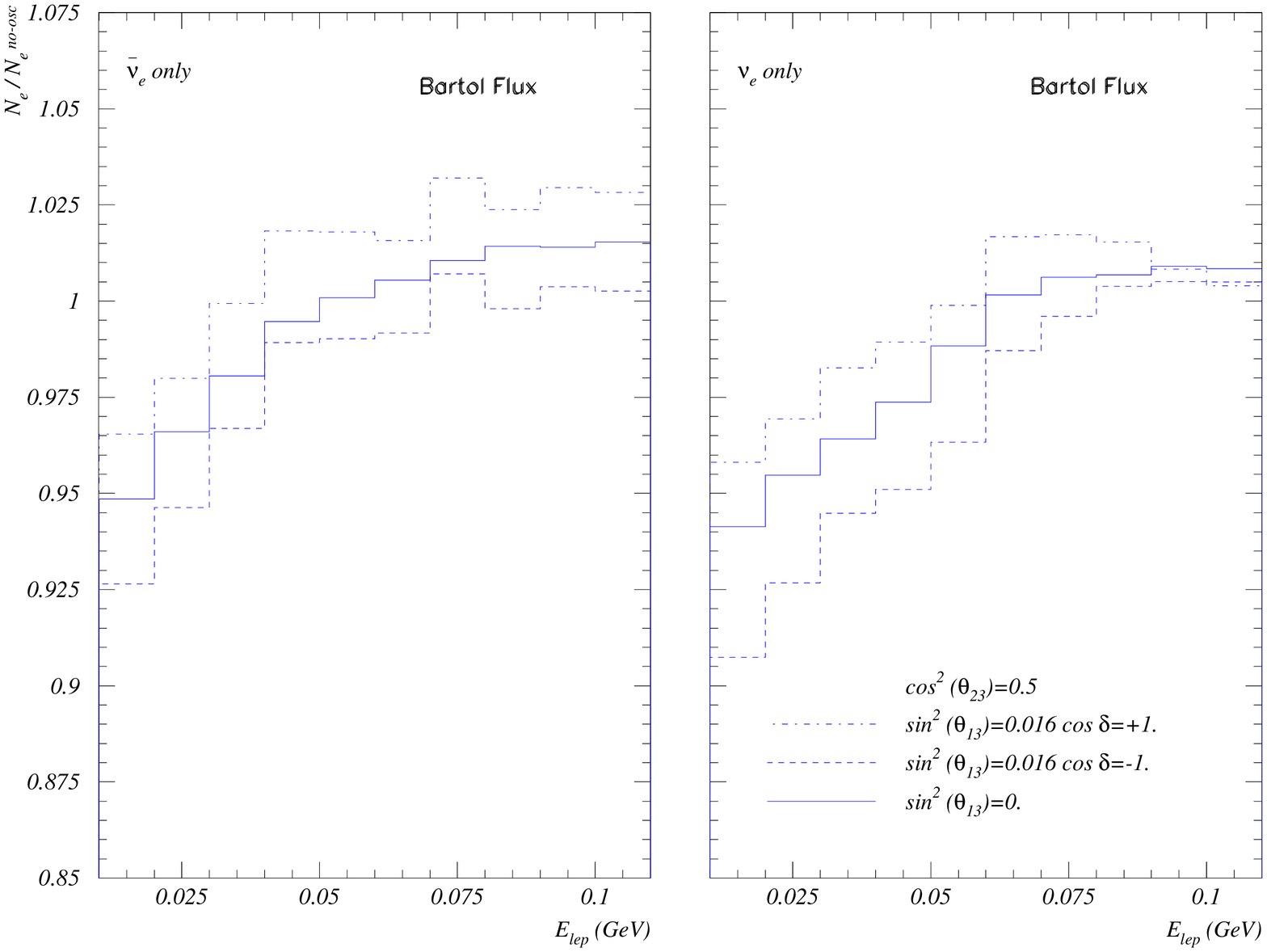}
\caption{Effect of the 1-3 mixing on the spectra of events induced by 
the electron neutrinos and antineutrinos.  
Shown are relative spectra of the $e-$like events for 
different values of 1-3 
mixing. We take $c_{23}^2 = 0.5$, $\sin^2 2\theta_{12}  = 0.82$, 
and $\Delta m^2_{21} = 7.3 \times 10^{-5}$ eV$^2$. 
{\it Left panel:} $\bar{\nu}_e$, {\it right panel:} ${\nu}_e$. 
}
\label{corr13}
\end{figure}

For the antineutrino channel $\sin^2 2\theta_m$ decreases with 
increase of energy. Estimations give 
$\langle P_2(E) \rangle = 0.089, 0.038, 0.013$ 
for $E = 20, 40, 80$ MeV correspondingly. 
For a given $c_{23}^2$ the change of $\epsilon_{\bar e}^{(0)}$ 
reflects this decrease of $\langle \bar{P}_2(E) \rangle$ and change of 
$\langle \bar{r}(E) \rangle$. In general, $\epsilon_{\bar e}^{(0)}$ 
decreases with energy. Distortion of spectrum can be characterized by 
$\Delta \epsilon_{\bar e}^{(0)} \equiv \epsilon_{\bar e}(100~ {\rm MeV}) 
- \epsilon_{\bar e}(10~ {\rm MeV})$. 
According to the figure the distortion $\Delta \epsilon_{\bar e}^{(0)} $   
is stronger for small $c_{23}^2$, it 
decreases from  $\sim  (8 - 9) \%$ for $c_{23}^2 = 0.35$, 
down to $\sim  (2 - 3) \%$ for $c_{23}^2 = 0.65$. 
Thus, if the shape of spectrum can be predicted with accuracy $\sim 1\%$, 
one can measure the deviation of  2-3 mixing from maximal 
by studying distortion of the spectrum in a  way 
which does not depend on uncertainties of the absolute value of 
neutrino flux. 
One possibility is to compare the integrated 
signal below and above $53$ MeV.  

For $c_{23}^2 >  0.6$ the dependence is not monotonous: 
with decrease of $E$,  $\epsilon_{\bar e}$ first increases and then decreases. 
For $c_{23}^2 >  0.6$ the oscillation effect is positive 
$\epsilon_{\bar e} > 0$ in whole energy range, whereas for 
$c_{23}^2 <  0.42$, $\epsilon_{\bar e}^{(0)} <  0$ everywhere,  
and for $c_{23}^2 \sim  0.5$ the correction $\epsilon_{\bar e}^0$ changes the sign.  

With change of $c_{23}^2$,  
$\Delta \epsilon_{\bar e}^{(0)} = \epsilon_{\bar e}(0.65) - \epsilon_{\bar e}(0.35) 
 = (11 - 12) \%$ for large  energies ($E \sim 90$ MeV), and 
$\Delta \epsilon_{\bar e} = 16 \%$ in the low energy bin. 
So,  if the absolute flux is known with accuracy few $\%$, the 2-3 mixing 
can be determined by just measuring the total number of events.  
Variation of the signal with $c_{23}^2$ is stronger 
here than in the sub-GeV sample  ($E > 100$ MeV),  although the number of events 
is smaller. 

In fig. \ref{f4} (right panel), we show  
the distortion of spectrum of the $e-$like 
events induced by scattering of $\nu_e$ on $^{16}O$. For neutrinos 
the flavor ratio $r$ is systematically smaller than for 
antineutrinos. Therefore the histograms shift to smaller values of 
ratio $N_e/N_e^0$. The suppression of  signal due to oscillations  
is stronger here. 
A character of the distortion is rather similar to that in 
the $\bar{\nu}-$case, and again, it can be traced from 
eqs.  (\ref{eps}, \ref{eps0}). Now $\langle P_2(E) \rangle$ changes 
with energy weaker. Indeed,  
$\kappa$ decreases,  whereas $\sin^2 2\theta_m$ increases  
with the energy increase and  the two changes partly compensate each other.  
For a given $c_{23}^2$ we obtain 
$\Delta \epsilon_e^{(0)} \equiv \epsilon_e^{(0)}(100~{\rm MeV}) - 
\epsilon_e^{(0)}(10~{\rm MeV}) \sim  6 \%$ for all  
values of $c_{23}^2$ in the interval $0.35 - 0.65$. 
For fixed energy, with change of $c_{23}^2$  one has   
$\epsilon_e^{(0)} (0.35) - \epsilon_e^{(0)} (0.65) 
\sim  15 \%$ for all energies.  
    
In fig. \ref{diff-sp} we compare the  distortions of spectra of events 
computed  with the neutrino fluxes published by different authors. This quantifies 
the present theoretical uncertainties.  
Fluxes from \cite{Honda:1995hz} (Honda) and \cite{Gaisser:1988ar}
(Bartol) lead to rather similar distortion 
although according to \cite{Honda:1995hz} $r(E)$ is flatter than in 
\cite{Gaisser:1988ar}  in the range $E > 60$ MeV, but below that 
energy $r(E)$  decreases sharper (see fig. \ref{compare3}). 
For large $c_{23}^2$ this leads to non-monotonous change of 
$\epsilon_{\bar e}$ with maximum at $E = (60 - 70)$ MeV, as we have marked 
before. 

Corrections due to  1-3 mixing described by 
$\epsilon_e^{(1)} + \epsilon_e^{(2)} + ... $  are shown in fig. 
\ref{corr13} for antineutrinos (left panel) and for neutrinos (right panel). 
All the features of the figures can be immediately understood 
using the analytic expressions derived above.  
Recall  that due to integration over the zenith angle  $I_2$ 
becomes negligible (see eq. (\ref{avPI})).  
Moreover, as we marked before, the corrections of the order $s_{13}^2$ are
additionally suppressed. Therefore in the first approximation
the corrections are given by  $\epsilon_e^{(1)}$ in  
eq. (\ref{eps1}). 
This term  weakly depends on $\theta_{23}$. $\langle R_{e\mu} \rangle$
can be estimated taking the constant density approximation: 
\be
\langle R_{e\mu} \rangle \approx \langle R_2 \rangle \sim 
-\frac{\kappa}{8} \sin 4 \theta^m_{12}.  
\ee
For directions below the horizon the corrections due to oscillations in atmosphere 
are additionally suppressed by $\cos^2 2 \theta_{12} \approx 0.18$  
(\ref{avcorr}). 

Consider first the effect for antineutrinos. In the lowest energy bin 
we have $\langle \bar{r} \rangle = 1.7$, $\kappa \sim 1.2$, and  
$\langle \bar{R}_{e\mu} \rangle  = - 0.085 \kappa$. 
Therefore the linear in $s_{13}$ correction equals 
\be 
\epsilon_{\bar e}^{(1)} (10~{\rm MeV})  \approx   
0.021 \left(\frac{s_{13} \cos\delta}{0.126} \right). 
\ee 
With increase of energy both $\langle \bar{R}_2 \rangle$ and 
$\langle \bar{r} \rangle$ increase and at $E = 100$ MeV 
($0.5 \sin 4\theta^m_{12} \approx 0.48$):  
\be
\epsilon_{\bar e}^{(1)} (100~{\rm MeV}) 
\approx
0.033 \left(\frac{s_{13} \cos\delta}{0.126}\right) 
\ee
in agreement with the results in  fig. \ref{corr13}. 
The corrections are CP-even being of the same sign for 
neutrinos and antineutrinos.  

For neutrinos we have $\langle r \rangle = 1.66$ in the lowest energy bin,   
and consequently, 
\be
\epsilon_{e}^{(1)} (10~{\rm MeV})
\approx
0.025 \left(\frac{s_{13} \cos\delta}{0.126}\right).
\ee
The correction decreases with energy 
due to decrease of $\langle {R}_{e\mu} \rangle$. 
For $E = 60$ MeV we have  $\langle r \rangle = 2.0$, 
$\langle {R}_{e\mu}  \rangle = - 0.035$, and therefore  
\be
\epsilon_{e}^{(1)} (60~{\rm MeV}) \approx
0.009  \left(\frac{s_{13} \cos\delta}{0.126}\right).
\ee
In the resonance $R_{e\mu} \approx 0$. Therefore, 
in the region around  $E \sim 90$ MeV the corrections due to 1-3 mixing 
are suppressed additionally, as can be seen from fig.~\ref{corr13}.  

Comparing results from different energy regions we can identify  
the 1-3 mixing effect (induced interference). 
Notice that corrections due to 1-3 mixing weakly depend on the 2-3 mixing which 
allows us to disentangle the effect of 1-3 mixing, 
$\epsilon_{e}^{(1)}$, and the effect due to the  1-2 mixing which is 
proportional to the deviation of 2-3 mixing from maximal. 

Corrections proportional to  $s_{13}^2$,  
\be 
\epsilon_{e}^{(2)} \sim 2s_{13}^2 [(\langle r \rangle s_{23}^2 - 1) + 
\langle\tilde{P}\rangle (2 - \langle r \rangle)], 
\ee 
are small: $< 0.005$ for $s_{13}^2 = 0.016$. 
Thus, the  degeneracy of $s_{13}^2$ and 
$c_{23}^2$ parameters can in principle be resolved 
using the energy dependence of the effect for neutrinos, however for 
antineutrinos the energy dependence is weak.

\section{Muon neutrino detection.  Invisible muon decays}

Muon neutrinos can be observed via the decay of invisible muons. 
Muons with energies below 
$160$ MeV do not produce signal in water Cherenkov detectors. 
They quickly lose energy, stop and decay at rest 
emitting electron (positron).    
In turn, most of these muons are produced in the detector 
in the quasi-elastic interactions  
$\nu_\mu + N \rightarrow N^\prime + \mu$. The effective (reconstructed) 
energy spectrum of these neutrinos has maximum 
at $0.16$ GeV and extends to  $0.25$ GeV \cite{Malek:2003ki}. 
In this energy range  $F_\mu \approx F_{\bar{\mu}}$, 
$F_e > F_{\bar{e}}$ and the flavor ratios equal $r = 2.03$, and  
$\bar{r} = 2.23$ \cite{Battistoni:2005pd}. 
Typical energies of neutrinos are high enough so that 
for estimations we can neglect the oscillations in atmosphere and 
use functions ${\bf D}$.  

Consider effects of oscillations on the number 
of electrons/positrons from invisible muon  
decay. Apparently,  the shape of energy spectrum 
of events is not influenced by oscillations and it has  
the  standard  form with maximum at $45$ MeV.     
\begin{figure}[htbp]
\includegraphics[scale=0.6]{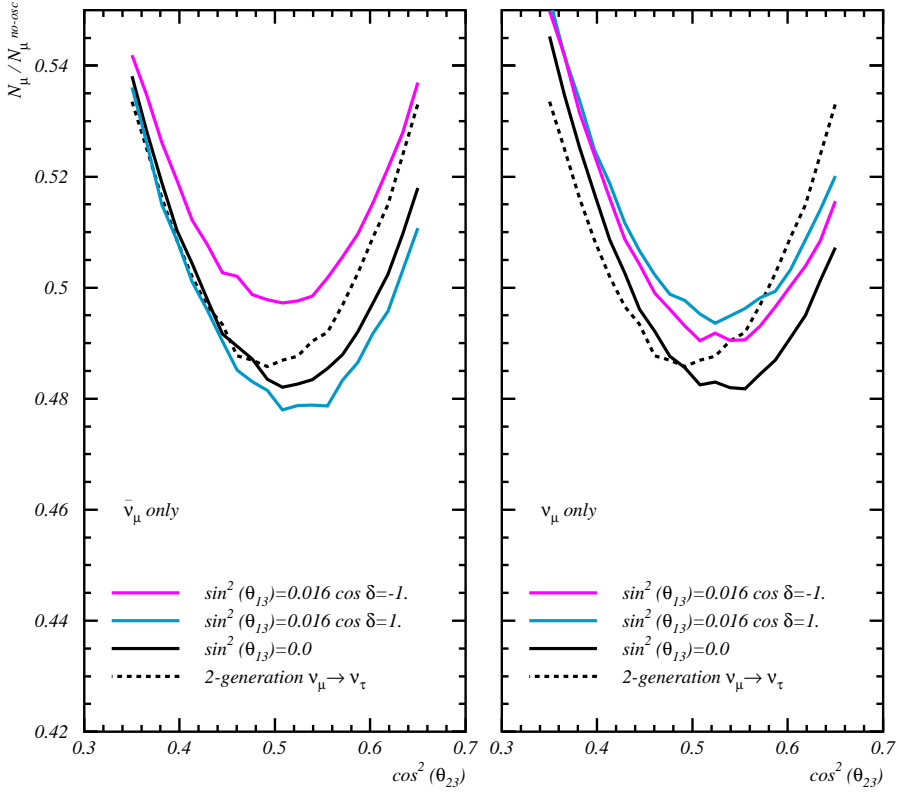}
\caption{Oscillation effects on the total number of invisible $\mu-$decay events  
induced by antineutrinos ({\it left panel}) 
and neutrinos ({\it right panel}).
Shown is the dependence of 
$N_{\mu}/N_{\mu}^0$ on $c_{23}^2$ for different values of 1-3 mixing. 
Dashed  line corresponds to the vacuum 
$\nu_\mu - \nu_\tau$ oscillations (zero 1-2 and 1-3 mixings).   
}
\label{mu-inv}
\end{figure}
In fig. \ref{mu-inv} we show the ratio of number of the invisible muon decays 
with and without oscillations,   
$N_{\mu}/N_{\mu}^0$, as function of $c_{23}^2$ for  
different values of $s_{13}$. 
All the features of this figure can be traced from the following 
semi-analytic consideration. According to (\ref{mumu-excess}) 
we have approximately
\be
\frac{N_{\mu}}{N_{\mu}^0} \sim 1 - \frac{1}{2}\sin^2 2\theta_{23} 
+ \epsilon_{\mu}^{(0)} + \epsilon_{\mu}^{(1)} + 
\epsilon_{\mu}^{(2)} + ... 
\label{expan-mu}
\ee
The zero order (in $s_{13}$) correction equals 
\be
\epsilon_{\mu}^{(0)} \approx - \langle {P_2} \rangle 
c_{23}^2 \left(c_{23}^2 - \frac{1}{\langle r \rangle}\right).  
\ee
The linear in $s_{13}$ corrections is given by 
\be
\epsilon_{\mu}^{(1)} \approx 
s_{13} \sin 2 \theta_{23}\cos\delta ~\langle R_2 \rangle
\left(2 c_{23}^2 - \frac{1}{\langle r \rangle} \right).
\label{13corr-mu}
\ee
The quadratic in $s_{13}$ corrections can be written as  
\beq
\epsilon_{\mu}^{(2)} & \approx & s_{13}^2 \left\{ \sin^2 2\theta_{23} 
\left[\cos^2 \delta \langle P_2 \rangle +   \frac{1}{2} 
\left\langle \frac{I^2_2 - R_2^2}{P_2} \right\rangle
 \right] - \right. 
\nonumber \\
& - & \left. 2 s_{23}^4 + \frac{1}{\langle r \rangle} 
\left(2 s_{23}^2 - \langle P_2 \rangle \right) \right\}. 
\label{mu-ev}
\eeq

In the constant density approximation we have 
\be
\frac{I^2_2 - R_2^2}{P_2}  =
1 - (1 + \cos^2 2 \theta_{12}^m) \sin ^2 \phi^m, 
\ee
and averaging over the zenith angle gives 
\be 
\left\langle \frac{I^2_2 - R_2^2}{P_2} \right\rangle = 
\frac{1}{2} \left(1  +  \frac{1}{2}\sin^2 2 \theta^m_{12}\right).   
\ee

The first term in (\ref{expan-mu}) is just the averaged vacuum $\nu_\mu - \nu_\tau$ 
oscillation probability. This probability  
(black curve) is symmetric with respect to maximal mixing,  
$c_{23}^2 = 0.5$. As expected, at this value of $c_{23}^2$ one has maximal 
suppression, 0.5. 

Corrections due to the oscillations driven by the 1-2 mixing, 
$\epsilon_{\mu}^{(0)}$, are asymmetric with respect to $c_{23}^2 = 0.5$.   
The corrections are zero at 
$c_{23}^2 = 1/\langle r \rangle$. They are positive at smaller 
$c_{23}^2$ and  negative at  $c_{23}^2 >  1/\langle r \rangle$. 
Maximal relative effect is   $ \approx 5\%$  suppression at 
$c_{23}^2 = 0.65$. The effect is weaker ($< 3\%$) for 
antineutrinos which is related to  
smaller probability $\bar{P}_2$ and larger value of $\bar{r}$   
only partly compensate the difference. 

The first order correction due to the 1-3 mixing, $\epsilon_{\mu}^{(1)}$, 
is given in (\ref{13corr-mu}). For the average energy of neutrinos 
producing invisible muons, $E = 160$ MeV, we have 
$\langle R_2 \rangle  = 0.04$ 
and $\langle \bar{R}_2 \rangle  = - 0.125$.
Notice that the  energies of neutrinos which generate the  
invisible muons are above the resonance, and therefore 
$R_2$ has opposite to $\bar{R}_2$ sign.  
Then for $c_{23}^2 = 0.65$ we obtain 
\be
\epsilon_{\mu}^{(1)} 
\approx 
0.004 \left(\frac{s_{13} \cos\delta}{0.126}\right).
\label{epsmu1}
\ee
For antineutrinos the corrections are much larger and of opposite sign: 
\be
\epsilon_{\bar{\mu}}^{(1)} 
\approx 
- 0.0127 \left(\frac{s_{13} \cos\delta}{0.126} \right).
\label{epsmubar1}
\ee
This is in agreement with the results of fig. \ref{mu-inv}. 

In contrast  to the direct $\nu_e$ - channel, 
here $s_{13}^2$ corrections are 
not suppressed by the screening factors and turn 
out to be of the same order as the 
linear corrections for not too small $s_{13}^2$.
Using (\ref{mu-ev}) we get for the quadratic term and $\cos\delta = 1$
\be
\epsilon_{\mu}^{(2)} \approx 
0.0085  \left(\frac{s_{13}^2 }{0.016}\right), 
\ee
and for antineutrinos the corresponding numerical coefficient equals 0.0054.  
The corrections are positive.  

Notice that the high energy neutrino fluxes responsible for 
the invisible muon production are affected by the solar activity  
weaker than fluxes at low energies.  At $E > 150$ MeV the difference of 
fluxes during minimum and maximum of solar activity is 
$(6 - 10) \%$, whereas at 30 MeV the difference  is about 24 \%. 
Variations at high energies are weaker. This can be used to 
disentangle the direct production and  signal 
from the invisible muon decays. 

\section{Future measurements}

High statistics of the sub-subGeV events can be 
obtained in future megaton-scale 
water Cherenkov detectors 
\cite{Jung:1999jq,Nakamura:2003hk,Mosca:2005mi,Autiero:2007zj}. 
In these detectors the signals of four different neutrino fluxes 
considered in the previous 
sections can be identified in the following way: 

1). The $\bar{\nu}_e$ quasi-elastic scattering on protons 
can be detected performing tagging of neutrons in 
correlation to the positron detection. In SuperKamiokande this will be possible 
with Gadolinium in the GADZOOK's version \cite{Beacom:2003nk}. 

2). The  $\bar{\nu}_\mu$ quasi-elastic scattering on protons has 
similar signatures: detection of neutron and positron from the muon 
decay. The difference from the previous case is that 
positrons appear in certain energy range and 
have known energy spectrum. So, the number of invisible decays can be 
extracted by fitting  the energy spectrum. 
(Also neutrons will have different energy characteristics.)

3). The electron neutrinos ${\nu}_e$'s are detected by their scattering on Oxygen.  
Electron appearing in this process should not be accompanied by 
neutron. (In principle, one can perform tagging of  nuclear transitions, 
e.g., detecting transition to exited state of nuclei.)

4). The ${\nu}_\mu-$ flux is detected via the invisible 
muon production on Oxygen. 
The standard decay spectrum of electrons and absence 
on neutron are the signatures of this interaction. 


Thus,  the antineutrino-induced events can be 
disentangled from neutrino interactions by the neutron tagging.  

Certain conclusions can be drawn from comparison of different signals. 
The oscillation corrections to the ${\nu}_e-$ signal are 
substantially enhanced in comparison to the $\bar{\nu}_e$ corrections 
if $c_{23}^2 < 0.5$. 
Comparison of the ${\nu}_e$ and $\bar{\nu}_e$ signals at $E > 60$ MeV 
can reveal the effect of 1-3 mixing.

\begin{figure}[htbp]
\includegraphics[scale=0.7]{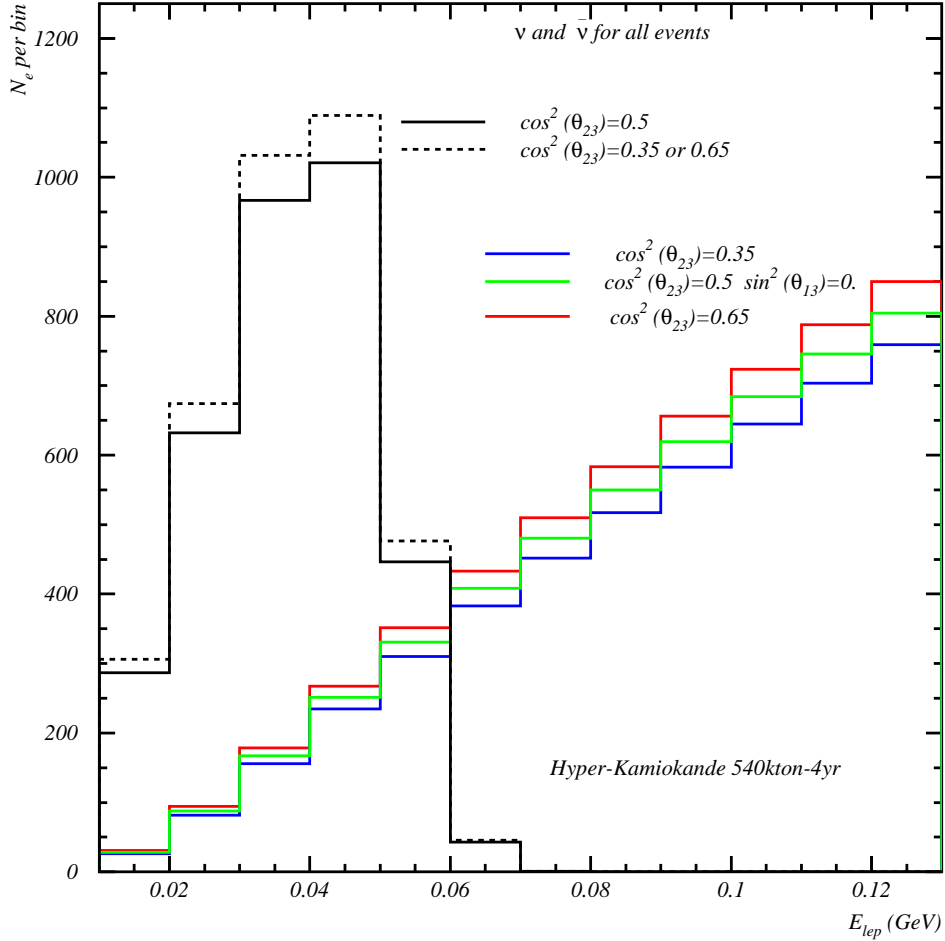}
\caption{The spectrum of neutrino and antineutrino events 
expected for 4 years of the  HyperKamiokande (0.5 Mton)  
data-taking.  
The red and blue histograms are for the e-like events 
directly produced by $\nu_e$  and $\bar{\nu}_e$ for two different 
values of 2-3 mixing. Black histograms correspond to the 
$e-$like events from the invisible 
muon decays computed respectively for $\cos^2 \theta_{23}=0.35$,
and $\cos^2 \theta_{23}=0.65$ (dashed lines) and 
$\cos^2 \theta_{23}=0.50$ (solid line). 
}
\label{f3}
\end{figure}

Fig. \ref{f3} illustrates dependence of the spectra of  the 
$e-$like events on the 2-3 mixing for $s_{13} = 0$. 
The histograms show spectra of the $e-$like events produced 
directly by the electron neutrinos and antineutrinos 
(red and blue), and from the invisible muon decay (black) 
for different values of $\cos \theta_{23}$. 
We show the number of $e$-like  events at HyperKamiokande 
(540 kton collected during 4 years).    
The difference of numbers of events for $c^2_{23}=0.35$ and  $c^2_{23}=0.65$  
in each bin is about $2\sigma$ (statistical error). 
Therefore using 9 bins one can distinguish 
two values of $c^2_{23}$ at $18 \sigma$ level.

\begin{figure}[htbp]
\includegraphics[scale=0.7]{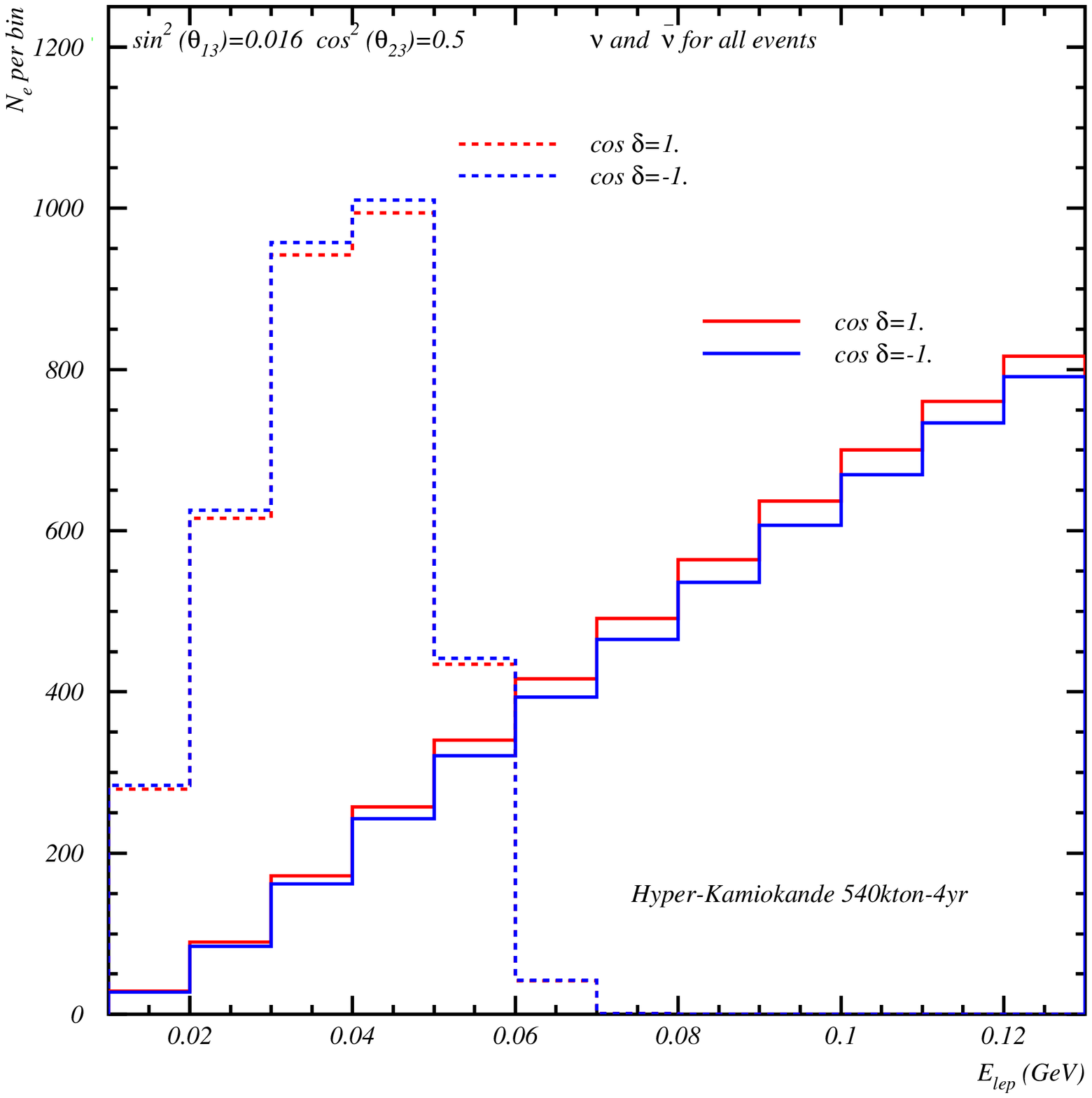}
\caption{Effect of the 1-3 mixing on the spectra of $e-$like events. 
The total number of $e-$like events induced by  ${\nu}_e$,  $\bar{\nu}_e$ 
and  invisible muons 
decays for the HyperKamiokande 4 years of data taking is given as a 
function of the charged lepton energy. Shown are the histograms
for $c_{23}^2 = 0.5$,  $s_{13}^2 = 0.016$ and 
$\cos \delta = + 1$ and $\cos \delta = - 1$.
}
\label{f11}
\end{figure}

The effect of 1-3 mixing is much stronger in the 
antineutrino signal than in the neutrino signal  
both for the electron  and muon neutrinos  
(i.e., for both components shown in the fig. \ref{f11}). 
The difference of histograms is essentially due to the linear 
$s_{13}$-corrections. 
Qualitatively the size of the effect can be immediately inferred from 
eq.  (\ref{eps1}) for direct component and from eqs. (\ref{13corr-mu},
\ref{epsmu1}, \ref{epsmubar1}) for the invisible 
muon decays.   

For maximal allowed value $s_{13}^2 = 0.03$ one would get 1.4 times 
larger difference of histograms than the one shown in fig. \ref{f11}. 

Notice that with increase of $c^2_{23}$ the number of invisible $\mu-$decays 
increases for $c^2_{23} > 0.5$ and decreases for $c^2_{23} < 0.5$. The number 
of e-like events  always increases with $c^2_{23}$. 
This can be used for measurements  of $c^2_{23}$.

Tagging of neutrons allows one to detect total signal 
produced by antineutrinos: $\bar{\nu}_e$ and $\bar{\nu}_\mu$. 
One can further disentangle the contributions 
from electron and muon antineutrinos by making the fit of 
spectral shape and using known spectrum of positrons from the 
muon decay at rest. This contribution is characterized by a single 
normalization parameter which can be then extracted from the data fit. 

If no neutron tagging occurs, one measures 
the total energy spectrum of electrons and positrons
(see figs. \ref{f3} and \ref{f11}). As follows from the figures, 
even in this case  one can extract certain information 
on deviation of the 2-3 mixing from maximal and on the 1-3 mixing. 

As we have mentioned before,  different components of spectrum have different 
dependence on the solar activity. This allows one, in principle, to disentangle the 
direct $\nu_e-$,  $\bar{\nu}_e-$signal,  and the signal from the invisible muon 
decays (muon neutrinos).

High statistics of future experiments will open a possibility 
to play with other characteristics such as  
angular distributions, directionality, time dependence.   
Also detection of the neutral current events can provide 
an important information. 

A very detailed study of the sub-subGeV events including 
determination of flavor and charge of the produced lepton 
will be possible with large volume liquid Ar detectors 
\cite{Cocco:2004ac,Cline:2006st},   
\cite{Ereditato:2005yx}.

\section{Conclusion}

1. There are several new features which appear at low energies 
($E \aprle 100$ MeV) 
in production, oscillations and detection of the atmospheric neutrinos. 
One of these features is substantial decrease of the flavor ratios 
with decrease of energy and their substantial deviation from 2 
in very low energy bins.  
As far as oscillations are concerned, there is strong averaging 
of oscillations due to integration over the angular variables. 
The oscillations in atmosphere become important. 
For events induced by the electron neutrinos the energy of electron (positron) 
is practically related to the energy of neutrino. 
In contrast, for  muon neutrinos detected via the decays of  invisible muons,  
information on the neutrino energy is practically lost. These events are induced,  
mainly,  by neutrinos in the energy interval (0.1 - 0.3) GeV. 

2. We performed both numerical and analytical study of the 
oscillations of sub-subGeV neutrinos. Dependence of the oscillation effects on 
$\sin \theta_{13}$, $\delta$ and deviation of the 2-3 mixing 
from maximal are given explicitly. For the rest one can use either numerical 
results or semi-analytical formulas. 

3. For neutrino trajectories below the horizon one should take into account 
the oscillations in two layer medium: the atmosphere and the Earth.   
We have presented the relevant oscillation probabilities 
and numbers of events as functions of $\sin \theta_{13}$, $\delta$, 
$\theta_{23}$ and in terms of  $P_2$, $R_2$, $I_2$. 
For the latter we give precise semi-analytical formulas using 
the Magnus expansion as well as  explicit expressions  
for the constant density. 

4. For the $e-$like events at  
$s_{13} = 0$ the corrections due to the oscillations driven by 
the 1-2 mixing can be as large as  $10 - 15 \%$ 
depending on  possible  deviations of the 2-3 mixing from maximal. The oscillations 
lead to change of the total number of events and also to 
distortion of the energy spectrum. The distortion is stronger 
in the low energy bins. The effect 
(change of the slope of the spectrum in the interval 10 - 100 MeV) 
can be as large as $(7 - 8)\%$.  

5. The 1-3 mixing leads to an additional, somehow smaller, effect. 
For maximal allowed values $s_{13}^2 \approx 0.03$, it can reach $\pm 4 \%$
at low energies, and $\pm 6 \%$ at high energies. 
The energy dependence of the effect is stronger for $\nu_e$.   
The linear corrections which dominate at low energies  vanish 
in the resonance $E \sim 100$ MeV. 
For antineutrinos corrections slightly decrease with energy. 
The corrections for antineutrinos are much larger than for neutrinos 
at $E \sim 100$ MeV.

6. In the  energy interval (0.1 - 0.3) GeV the $\nu_{\mu}$-flux 
is detected via the decay of invisible muons. 
The total number of $e-$like events from these muons  depends on 
the 2-3 mixing in the lowest order. 
For allowed values of the deviation of the 2-3 mixing 
from maximal the effect  can be as large as $5\%$. Inclusion of the 1-2 mixing 
gives (for $s_{13} = 0$) maximum $5 \%$ effect for neutrinos 
and  $3 \%$ effect for antineutrinos. This maximal value is realized at 
$c_{23}^2 = 0.65$. For $c_{23}^2 < 0.5$ the effect is smaller. 

The 1-3 mixing  produces comparable to 1-2 mixing corrections. 
The largest effect is for 
antineutrinos  and $c_{23}^2 = 0.65$:  $\epsilon_\mu^{(1)} \sim 0.02$. 
The linear in $s_{13}$ correction  is strongly suppressed for neutrinos. 
The quadratic corrections are not negligible and can be of the order 
of $0.01$. 

7. There is strong degeneracy of parameters $s_{13} \cos \delta$  
and $c_{23}^2$, especially for the invisible decays.  
Inclusion of the 1-3 mixing effects only slightly enlarge the region of 
possible values of ratio $N_\mu/N_\mu^0$. 

8. The signals from interactions of 4 different fluxes $\nu_e$, $\nu_\mu$
$\bar{\nu}_e$, $\bar{\nu}_\mu$ can be disentangled by 
tagging the accompanying neutrons,  
studying the shape of energy spectrum as well as the time dependence of 
signals.  
Confronting events of different types as well as 
events at low and high energies allows one to reduce the degeneracy 
of oscillation parameters. 

9. The number of presently detected events 
(at SuperKamiokande I) due to the interactions of very low 
energy neutrinos is about 
few hundreds which provides an accuracy  
of determination of the oscillation parameters not better than $10 \%$,   
and an additional problem is uncertainties of neutrino fluxes. 
So,  to study oscillation effects discussed in this paper one needs 
much larger statistics which can be achieved with the 
Megaton-scale detector.

10.  The low energy atmospheric neutrinos can be used to measure 
deviation of 2-3 mixing from maximal, the 1-3 mixing and the phase $\delta$. 
They can be used to search for new physics. Apparently for the 
latter knowledge of the standard oscillation effects computed in this 
paper is necessary. Understanding of fluxes of these neutrinos is also important 
for future studies 
of the relic supernova neutrinos. 

\section*{Acknowledgments}

The authors are grateful to G. Battistoni and T. Stanev for 
discussions of the atmospheric neutrino fluxes at low energies. 
O. L.G. P. thanks FAPESP, CNPQ and ICTP for support.

\section*{Appendix A: Probabilities $P_2$, $R_2$ and $I_2$ in the second order 
of the Magnus expansion. }

The probability functions ${\bf D}$ in the second order Magnus expansion 
can be found using definitions (\ref{p-r-i}) and results of \cite{ara9}:   
\beq
P_2 & = & 
\left[\left(\cos I_{tot} \sin \phi^{ad}
-  \frac{I_{\theta \theta}}{I_{tot}} \sin I_{tot} \cos \phi^{ad}\right) 
\sin 2\theta_m^0 + \frac{I_\theta}{I_{tot}} \sin I_{tot} \cos 2\theta_m^0 
\right]^2, 
\nonumber\\
R_2 & = &- \frac{1}{2} \sin 4\theta_m^0
\left[\left(\cos I_{tot} \sin \phi^{ad}
- \frac{I_{\theta \theta}}{I_{tot}}
\sin I_{tot} \cos\phi^{ad}\right)^2
- \left(\frac{I_\theta}{I_{tot}}\right)^2 \sin^2 I_{tot}\right]-  \nonumber\\
 & - & \cos 4\theta_m^0 \left(\cos I_{tot} \sin \phi^{ad} - \frac{I_{\theta 
\theta}}{I_{tot}}
\sin I_{tot} \cos\phi^{ad}\right) \frac{I_\theta}{I_{tot}} \sin I_{tot}, 
\nonumber\\
I_2 & = & \frac{1}{2} \cos^2 I_{tot} \sin 2\theta_m^0 \sin 2\phi^{ad}
+ \frac{I_\theta}{2 I_{tot}} \sin 2 I_{tot} \cos 2\theta_m^0 
\cos \phi^{ad} + \nonumber\\
& + & I_{\theta \theta} \left[ 
- \frac{\sin 2 I_\theta}{2 I_{tot}}  \sin 2\theta_m^0
\cos 2\phi^{ad}
- \frac{I_{\theta \theta}}{2I_{tot}^2}
\sin^2 I_{tot}\sin 2\theta_m^0 \sin 2\phi^{ad} + \right.
\nonumber\\ 
& + & \left. I_\theta \frac{\sin^2 I_\theta}{I_{tot}^2} 
\cos 2\theta_m^0 \sin \phi^{ad}
\right]. 
\eeq
Here 
\be
I_{tot} \equiv  \sqrt{I_\theta^2 + I_{\theta \theta}^2}, 
\nonumber
\ee
and 
\be
I_{\theta \theta} =  4 \int^{x_f}_{\bar{x}} dx 
\int^{x}_{\bar{x}} dy  \left[\frac{d \theta_m (x)}{dx}\right]
\left[\frac{d \theta_m (y)}{dy}\right]
\sin \phi^{ad} (x) \cos\phi^{ad} (y) .
\nonumber
\ee

\section*{Appendix B: Functions $\tilde{\bf D}$ for two layers.}  

According to eqs. (\ref{s-tot}, \ref{matr2}, \ref{matrA})
the elements of $S$-matrix in the two layer case equal 
\begin{eqnarray}
\tilde{A}_{ee} & = & A_{ee}^{\prime} (c_{\phi} + i \cos 2\theta_{12} s_\phi) - 
i A_{e\mu}^{\prime}\sin 2\theta_{12} s_\phi ,  
\nonumber \\
\tilde{A}_{e\mu} & = & A_{e\mu}^{\prime} (c_{\phi} - i \cos 2\theta_{12} s_\phi) - 
i A_{ee}^{\prime}\sin 2\theta_{12} s_\phi , 
\nonumber \\
\tilde{A}_{\mu e} & = & A_{e\mu}^{\prime} (c_{\phi} + i \cos 2\theta_{12} s_\phi) - 
i A_{\mu \mu}^{\prime}\sin 2\theta_{12} s_\phi , 
\nonumber \\
\tilde{A}_{\mu \mu} & = & A_{\mu \mu}^{\prime} 
(c_{\phi} - i \cos 2\theta_{12} s_\phi) - 
i A_{e\mu}^{\prime}\sin 2\theta_{12} s_\phi , 
\label{tilde-a}
\end{eqnarray}
and $\tilde{A}_{33} = exp[-i2(\phi_3^m + \phi_3)]$, where 
$\phi_3^m$ is given in (\ref{phase}) and $\phi_3$ 
is the phase acquired in atmosphere. 
We present here 
the probability functions  $\tilde{\bf D}$ defined as 
\be 
\tilde{A}_{e \mu }^* \tilde{A}_{ee} \equiv R_{e\mu} + i I_{e\mu}, ~~~~
\tilde{A}_{\mu e}^* \tilde{A}_{ee} \equiv R_{\mu e} + i I_{\mu e} 
\label{deffi}
\ee
(see also (\ref{5funct})).   
Using expressions for the amplitudes (\ref{tilde-a}) we obtain 
\be
\tilde{P} = P_2   
+ R_2 \sin 4 \theta_{12} \sin^2 \phi 
+ I_2 \sin 2\theta_{12} \sin 2\phi 
+ (1 - 2 P_2) \sin^2 2\theta_{12} \sin^2 \phi, 
\nonumber
\ee
\be
R_{e \mu} = R_2(1 - 2 \cos^2 2\theta_{12} \sin^2 \phi) 
- I_2 \cos 2\theta_{12} \sin 2\phi 
- \left(\frac{1}{2} - P_2\right) \sin 4 \theta_{12} \sin^2 \phi, 
\nonumber
\ee
\be
I_{e \mu} = I_2(1 - 2 \sin^2 \phi) 
+ R_2 \cos 2\theta_{12} \sin 2\phi 
+ \left(\frac{1}{2} - P_2\right) \sin 2\theta_{12} \sin 2\phi, 
\nonumber
\ee
\be
R_{\mu e} = R_2(1 - 2 \sin^2 2\theta_{12} \sin^2 \phi) 
+  \frac{R_2 I_2}{P_2}\sin 2\theta_{12} \sin 2\phi 
- \left(\frac{1}{2} - \frac{R_2^2}{P_2}\right) \sin 4\theta_{12} \sin^2 \phi, 
\nonumber
\ee
\be
I_{\mu e} = I_2(1 - 2 \sin^2 2\theta_{12} \sin^2 \phi) 
+ \frac{R_2 I_2}{P_2} \sin 4\theta_{12} \sin^2 \phi 
+ \left(\frac{1}{2} - \frac{R_2^2}{P_2}\right)\sin 2\theta_{12} \sin2 \phi.  
\ee
The limit $P_2 = R_2 = I_2 = 0$ corresponds to 
oscillations in the atmosphere only (trajectories above the horizon). 
In this case the probabilities are given by the last terms in each 
expression which agree with one layer results (\ref{prob-con}). 


\section*{Appendix C: Constant density case}

In the constant density approximation the evolution matrix 
in matter is given by 
\be  
S^\prime \approx 
\left(\begin{array}{cc} 
c_{\phi}^m + i \cos 2\theta^m_{12} s_\phi^m  & - i \sin 2\theta^m_{12} s_\phi^m    
\\
- i \sin 2\theta^m_{12} s_\phi^m   &    c_{\phi}^m - i \cos 2\theta^m_{12} 
s_\phi^m  
\end{array}
\right) ~,  
\label{matrAc}
\ee
where $s_\phi^m \equiv \sin \phi^m$, $c_\phi^m \equiv \cos \phi^m$,
and $\phi^m$ is the oscillation phase in matter. 
The total evolution matrix in two layers 
(vacuum and matter) is $S^{tot} = S^\prime S_A$. 
Then in the basis $\nu^{\prime}$  the amplitudes for  two layers,  
$A_{\alpha \beta} =  [S^{tot}]_{\alpha \beta}$,  
equal 
\beq 
\tilde{A}_{ee} & = & c_\phi^m c_\phi - \cos2(\theta^m_{12} - \theta_{12})
s_\phi^m s_\phi + i (\cos 2\theta_{12} s_\phi c_\phi^m 
+ \cos2 \theta^m_{12} s_\phi^m c_\phi).
\nonumber\\ 
\tilde{A}_{e\mu} & = &  
\sin 2(\theta_{12} - \theta^m_{12}) s_\phi^m s_\phi
-  i (\sin 2\theta_{12} s_\phi c_\phi^m 
+    \sin 2\theta^m_{12} s_\phi^m c_\phi),\\
\tilde{A}_{\mu \mu} & = & \tilde{A}_{ee}^*, ~~~~
\tilde{A}_{\mu e} = - \tilde{A}_{e\mu}^*. 
\nonumber
\eeq  

The functions $\tilde{\bf D}$ (\ref{deffi}) can be  presented in the following form
\beq
\tilde{P}^c & = & 
\sin^2 2\theta^m_{12} \sin^2 \phi^m \cos^2 \phi  
+ \sin^2 2\theta_{12} \sin^2 \phi \cos^2 \phi^m + \\
\nonumber
& + & \frac{1}{2} \sin 2\theta^m_{12} 
\sin 2\theta_{12} \sin 2\phi^m \sin 2\phi + \sin^2 2(\theta^m_{12} -\theta_{12}) 
\sin^2 \phi^m \sin^2 \phi,  
\eeq
\beq
2 R_{e \mu}^c & = &  - \sin 4\theta^m_{12} \sin^2 \phi^m \cos^2 \phi  
- \sin 4\theta_{12} \sin^2 \phi \cos^2 \phi^m -   
\nonumber\\
 & - & \sin 2\theta^m_{12} \cos 2\theta_{12} \sin 2\phi^m \sin 2\phi 
+  \sin 4(\theta^m_{12} -\theta_{12}) \sin^2 \phi^m \sin^2 \phi 
\eeq
\beq
2 I_{e \mu}^c & = &  
\sin 2\theta^m_{12} \sin 2\phi^m \cos^2 \phi  
+ \sin 2\theta_{12} \sin 2\phi \cos^2 \phi^m - 
\nonumber\\
& - & \sin (4\theta^m_{12} - 2 \theta_{12}) \sin 2\phi \sin^2 \phi^m 
- \sin 2\theta^m_{12}  \sin^2 \phi \sin 2 \phi^m ,
\eeq
\beq
2 R_{\mu e}^c & = &  
- \sin 4\theta^m_{12} \sin^2 \phi^m \cos^2 \phi  
- \sin 4\theta_{12} \sin^2 \phi \cos^2 \phi^m -
\nonumber\\
& - & \sin 2\theta_{12} \cos 2\theta^m_{12} \sin 2\phi^m \sin 2\phi 
- \sin 4(\theta_{12}^m -\theta_{12}) \sin^2 \phi^m \sin^2 \phi , 
\eeq
\beq
2 I_{\mu e}^c & = &  
\sin 2\theta_{12}^m \sin 2\phi^m \cos^2 \phi  
+ \sin 2\theta_{12} \sin 2 \phi \cos^2 \phi^m -
\nonumber\\
& - & \sin (4\theta_{12} - 2 \theta^m_{12}) \sin^2 \phi  \sin 2\phi^m 
- \sin 2\theta_{12} \sin 2\phi \sin^2 \phi^m.  
\eeq
Notice that in the limit $\phi \rightarrow 0$ the first term 
in each expression corresponds to 
the probability of oscillations in matter, the second term 
gives the probability in vacuum ($\phi_m \rightarrow 0$). 
These two terms are related by the interchange of the  
vacuum and matter characteristics: 
$\theta_m \leftrightarrow \theta$, $\phi_m \leftrightarrow \phi$. 
The rest is the interference of the 
oscillation effects in the two layers. 
The first two terms are the same in 
$R_{\mu e}$ and $R_{e \mu}$ as well as in 
$I_{\mu e}$ and $I_{e \mu}$; in turn, these probabilities differ by the 
interference terms. 
The probability $\tilde{P}$ is symmetric with respect to 
interchange of the matter and vacuum characteristics.

For our discussion it is convenient to write the probabilities  in the 
form given in eq. (\ref{form+}):   
\beq
\tilde{P}^c & = & 
\sin^2 2\theta^m_{12} \sin^2 (\phi^m + \phi)
- (\sin^2 2\theta_{12}^m - \sin^2 2\theta_{12})
\sin^2 \phi \cos^2\phi^m - 
\nonumber \\
& - & \frac{1}{2}  \sin 2\theta_{12}^m (\sin 2\theta_{12}^m - \sin 2\theta_{12}) 
\sin 2\phi \sin 2\phi^m 
+ \sin^2(\theta_{12}^m - \theta_{12}) \sin^2\phi \sin^2\phi^m, 
\nonumber\\
R_{e \mu}^c & = &  
- \frac{1}{2} \sin 4\theta^m_{12} \sin^2 (\phi^m + \phi)  
- \frac{1}{2} \sin 2\theta^m_{12} (\cos 2\theta_{12} - \cos 2\theta^m_{12})
\sin 2\phi \sin 2\phi^m + 
\nonumber \\
& + & \frac{1}{2}\sin^2\phi \left[(\sin 4\theta^m_{12} - \sin 4\theta_{12}) 
\cos^2\phi^m
+ \sin 4(\theta^m_{12} - \theta_{12}) \sin^2\phi^m \right] ,  \nonumber\\
R_{\mu e}^c & = &
- \frac{1}{2} \sin 4\theta^m_{12} \sin^2 (\phi^m + \phi)
- \frac{1}{2} \cos 2\theta^m_{12} (\sin 2\theta_{12} - \sin 2\theta^m_{12})
\sin 2\phi \sin 2\phi^m + 
\nonumber \\
& + & \frac{1}{2}\sin^2\phi \left[(\sin 4\theta^m_{12} - \sin 4\theta_{12})
\cos^2\phi^m
- \sin 4(\theta^m_{12} - \theta_{12}) \sin^2\phi^m \right] ,  \nonumber\\
I_{e \mu}^c & = &  
\frac{1}{2} \sin 2\theta^m_{12}  \sin 2(\phi^m + \phi)  
+ \frac{1}{2}\sin 2\phi 
\left[(\sin 2\theta_{12} - \sin 2\theta^m_{12}) \cos^2\phi^m + \right.
\nonumber \\
& + & \left. (\sin 2\theta_{12}^m - \sin(4\theta^m_{12} 
- 2\theta_{12})) \sin^2 \phi^m \right],  
\nonumber\\
I_{\mu e}^c & = &
\frac{1}{2} \sin 2\theta^m_{12}  \sin 2(\phi^m + \phi)
+ \frac{1}{2}
(\sin 2\theta_{12} - \sin 2\theta^m_{12}) \sin 2\phi \cos 2\phi^m + 
\nonumber \\
& + & \frac{1}{2}\left[\sin 2\theta_{12}^m - \sin(4\theta_{12}
- 2\theta_{12}^m)\right] \sin^2 \phi \sin 2 \phi^m . 
\label{pri-const}
\eeq



\providecommand{\href}[2]{#2}\begingroup\raggedright\endgroup

\end{document}